\documentclass[journal,twoside,web]{ieeecolor2}
\usepackage{generic}
\usepackage{cite}
\usepackage{amsmath,amssymb,amsfonts}
\usepackage{algorithmic}
\usepackage{graphicx}
\usepackage{textcomp}

\usepackage{siunitx}
\usepackage{xspace}
\usepackage[normalem]{ulem}
\usepackage{makecell}
\usepackage{booktabs}
\usepackage[english]{babel}
\usepackage{paralist}
\usepackage{subcaption}
\usepackage{caption}
\usepackage{multirow}
\usepackage{array}
\usepackage[export]{adjustbox}
\usepackage{comment}

\makeatletter
\DeclareRobustCommand\onedot{\futurelet\@let@token\@onedot}
\def\@onedot{\ifx\@let@token.\else.\null\fi\xspace}
\def\eg{\emph{e.g}\onedot} 
\def\ie{\emph{i.e}\onedot}

\def\etal{\emph{et al}\onedot}

\makeatother
\def\BibTeX{{\rm B\kern-.05em{\sc i\kern-.025em b}\kern-.08em
    T\kern-.1667em\lower.7ex\hbox{E}\kern-.125emX}}
\begin{document}
\title{Towards Seamless Integration of Magnetic Tracking into Fluoroscopy-guided Interventions}
\author{Shuwei Xing, Mateen Mirzaei, Wenyao Xia, Inaara Ahmed-Fazal, Utsav Pardasani, Uditha Jarayathne, Scott Illsley, Leandro Cardarelli Leite, Aaron Fenster, \IEEEmembership{Life Fellow, IEEE}, Terry M. Peters, \IEEEmembership{Life Fellow, IEEE} and Elvis C. S. Chen, \IEEEmembership{Member, IEEE}
\thanks{This work was supported in part by the Canadian Institutes of Health Research Foundation Grant 201409, in part by the Natural Sciences and Engineering Research Council of Canada under Grant 2024-06674,in part by the Canadian Foundation for Innovation under Grant 36199, in part by INOVAIT, in part by Canon Medical Systems. (Corresponding author: Shuwei Xing) }
\thanks{Shuwei Xing, Mateen Mirzaei, Wenyao Xia, and Inaara Ahmed-Fazal are with the Robarts Research Institute, Western University, London, ON N6A, 3K7, Canada}
\thanks{Utsav Pardasani, Uditha Jarayathne and Scott Illsley are with the Northern Digital Inc., Waterloo, ON, N2V 1C5, Canada}
\thanks{Leandro Cardarelli Leite is with the Department of Medical Imaging, Western University, London, ON N6A 3K7, Canada}
\thanks{Aaron Fenster, Terry M. Peters and Elvis C.S. Chen are with the Robarts Research Institute, the School of Biomedical Engineering, and the Department of Medical Biophysics, Western University, London, ON N6A 3K7, Canada}}

\maketitle

\begin{abstract}
The 2D projective nature of X-ray radiography presents significant limitations in fluoroscopy-guided interventions, particularly the loss of depth perception and prolonged radiation exposure. Integrating magnetic trackers into these workflows is promising; however, it remains challenging and under-explored in current research and practice. To address this, we employed a radiolucent magnetic field generator (FG) prototype as a foundational step towards seamless magnetic tracking (MT) integration. A two-layer FG mounting frame was designed for compatibility with various C-arm X-ray systems, ensuring smooth installation and optimal tracking accuracy. To overcome technical challenges, including accurate C-arm pose estimation, robust fluoro-CT registration, and 3D navigation, we proposed the incorporation of external aluminum fiducials without disrupting conventional workflows. Experimental evaluation showed no clinically significant impact of the aluminum fiducials and the C-arm on MT accuracy. Our fluoro-CT registration demonstrated high accuracy (mean projection distance $\approx$ \SI{0.7}{\milli\metre}), robustness (wide capture range), and generalizability across local and public datasets. In a phantom targeting experiment, needle insertion error was between \SI{2}{\milli\metre} and \SI{3}{\milli\metre}, with real-time guidance using enhanced 2D and 3D navigation. Overall, our results demonstrated the efficacy and clinical applicability of the MT-assisted approach. To the best of our knowledge, this is the first study to integrate a radiolucent FG into a fluoroscopy-guided workflow.
\end{abstract}

\begin{IEEEkeywords}
Fluoroscopy-guided interventions, magnetic tracking, radiolucent field generator, 2D-3D registration, target registration error.
\end{IEEEkeywords}

\section{Introduction}\label{introduction}
%fluoroscopy for endovascular aortic repair~\cite{BAK+2018}
%cardiovascular~\cite{HBB+2004}
%fluoroscopy for pedical screw~\cite{GPP+2012}
%fluoroscopy for pain management~\cite{Wang2018}
%The Role of Intraoperative Navigation in Orthopaedic Surgery~\cite{KMG+2019}

\IEEEPARstart{X}{-ray} fluoroscopy continues to be the predominant modality for intra-operative image guidance, ubiquitously employed across various domains including cardiovascular, endovascular, orthopedic, and neuro-interventions, as well as in pain management and biopsies\cite{merloz2007fluoroscopy, cazzato2020spinal, nijkamp2019prospective}. Fluoroscopy provides high-contrast visualization of osseous structures, contrast-enhanced regions, and metallic objects, thereby enabling interventionists to monitor and safeguard the real-time advancement of surgical instruments. To appreciate the instrument depth, the C-arm must be oriented perpendicular to the instrument’s trajectory and, after evaluating the depth, reverted to its original pose to monitor the instrument's orientation. Additionally, the C-arm requires efficient positioning to achieve desired radiographic views for optimizing instrument placement, a process that often necessitates repeated image acquisitions, known as “fluoro-hunting”. The requirement to reposition the C-arm and perform repeated image acquisitions stems from the 2D projective nature of X-ray radiography. This nature contributes to the main limitations of fluoroscopy-guided interventions, notably the loss of depth perception and prolonged radiation exposure for both patients and medical personnel. 

To mitigate these limitations, one promising strategy is to integrate spatial trackers into fluoroscopy-guided surgical workflows. Specifically, pose sensors are attached to surgical instruments, and optionally to the patient’s anatomy and the C-arm, each of which is spatially calibrated and co-registered with one another. The co-registration of all tracked objects into a common coordinate system enables the development of surgical navigation systems, as well as serving as the technological foundation for image fusion and 3D visualization. Merloz~\etal~\cite{merloz2007fluoroscopy} incorporated an optical tracker into fluoroscopy-guided pedicle screw insertion, marking one of the earliest clinical evaluations of computer-assisted spine surgeries. Their study demonstrated that the navigated approach reduced the rate of cortex penetration and radiation exposure time, but at the expense of a longer operative time. Recently, Gao~\etal~\cite{gao2022fluoroscopy} developed a fluoroscopy-guided robotic system designed for transforaminal lumbar epidural injections, incorporating an optical tracker to track both an injection device and a surgical robot. Similarly, Bakhtiarinejad~\etal~\cite{bakhtiarinejad2023surgical} proposed a fluoroscopy guidance system for osteoporotic hip augmentation, further expanding the application of the optical tracking-based solution. However, due to their reliance on line-of-sight, optical trackers pose practical challenges for optimizing their spatial configuration within the operating room to ensure simultaneous tracking of surgical instruments, patient anatomy, the C-arm, as well as some potential external hardware such as a surgical robot. Furthermore, optical trackers are incapable of tracking flexible surgical instruments, such as catheters, guidewires, and flexible endoscopes seated subcutaneously~\cite{ramadani2022survey}. Consequently, the utilization of optical trackers is often restricted to neurosurgery~\cite{mccutcheon2004frameless} and orthopedics, where the surgical target is rigid and exposed. 

Magnetic tracking (MT) has gained considerable attention due to its capability to track flexible instruments placed subcutaneously without the constraints of line-of-sight. Recently, Ramadani~\etal~\cite{ramadani2022survey} conducted a thorough review of tracking techniques for curvilinear catheters in interventional radiology and endovascular interventions, highlighting the efficacy of MT in these contexts. Nonetheless, it is well-recognized that MT is susceptible to interference from ferromagnetic materials. Over two decades ago, Hummel~\etal~\cite{hummel2002evaluation} demonstrated that X-ray fluoroscopy settings could cause distortion of magnetic fields, thereby affecting the tracking robustness of MT. More recently, advancements in MT technology have been explored. Yaniv~\etal~\cite{yaniv2006fluoroscopy} investigated the effect of bi-plane fluoroscopes on the accuracy of the Aurora MT system (Northern Digital Inc., Ontario, Canada). Their findings indicated that the presence of bi-plane fluoroscopes in a clinical setup has a negligible effect on the tracking accuracy, with a maximal error of \SI{1.4}{\milli\metre}. Similarly, Lugez~\etal~\cite{lugez2015electromagnetic} demonstrated that MT has the potential to provide effective assistance during surgical and interventional procedures. Additionally, Xu~\etal~\cite{xu2023surgical} employed a magnetic tracker for pedicle screw insertion and compared it with a robot-assisted approach, demonstrating the efficacy of MT-tracked over the robot-assisted approaches in terms of faster instrument placement and reduced X-ray exposure. 

It is important to note that standard magnetic field generators (FGs) are constructed with radiopaque materials, which introduce metal-induced imaging artifacts in radiographs, potentially limiting their clinical utility. To mitigate this issue, Yoo~\etal~\cite{yoo2013electromagnetic} incorporated a window FG, in which the FG coils are configured around the periphery of an open aperture, into the surgical table. While the FG window design effectively eliminated imaging artifacts induced by metal coils, it resulted in diminished tracking accuracy and restricted the use of fluoroscopy to small-oblique views. Recent advancements include radiolucent FGs that minimize metal-induced imaging artifacts under X-ray fluoroscopy~\cite{xia2023x, o2021radiolucent}, as well as error compensation techniques developed to mitigate the adverse effects of ferromagnetic interference, such as distorting the magnetic tracking field and limiting the effective tracking range. However, a solution that seamlessly integrates MT into fluoroscopy-guided surgical workflows remains challenging, particularly with respect to the latest radiolucent FGs. In this study, we focus on addressing the following unmet clinical needs to improve the fluoroscopy-guided procedure:
\begin{enumerate}
    \item \textit{The ability to enable robust 3D navigation}. Recent studies~\cite{stockle2007image} demonstrate that 3D navigation is more effective and clinically preferred compared to conventional 2D fluoroscopy guidance. It offers the advantages of 2D and 3D reformations in CT-like quality, thereby addressing the limitation of depth perception inherent in 2D fluoroscopy, along with 3D tracking of surgical instruments. One critical technical component for achieving 3D navigation is to effectively register fluoroscopic images with pre-operative CTs~\cite{cleary2010image}. Despite fluoro-CT registration methods having been extensively investigated over the past decades, this technique has not yet become a standard component in fluoroscopy-guided interventions~\cite{unberath2021impact}. The widespread adoption of this approach has been restrained by challenges, including limited capture range and sensitivity to the initial pose estimate~\cite{markelj2012review, esteban2019towards}. Therefore, an easy-to-use and robust fluoro-CT registration solution remains a clinical need to benefit 3D navigation.

    \item \textit{Compatibility with standard surgical workflows}. To ensure optimal MT accuracy, the FG must be placed in proximity to the surgical or therapeutic site~\cite{franz2014electromagnetic}. There are several configurations for placing the FG, including above the patient~\cite{franz2014electromagnetic}, on the surgical table alongside the patient~\cite{lugez2015electromagnetic}, or adjacent to the patient but off the surgical table~\cite{yaniv2006fluoroscopy}. Additionally, the FG setup must not obstruct the fluoroscopic view and should minimize disruption to the standard surgical workflow. However, these current configurations pose challenges for seamless integration into various fluoroscopy-guided procedures.

In addition, in the context of fluoroscopy-guided procedures, two types of C-arm imaging systems are generally used: manually driven and mechanically driven. The more commonly used former approach requires manual maneuvering to achieve desired views. Typically, its intrinsic parameters, such as pixel spacing, image dimension and source-to-detector (SID) distance, remain fixed. Mechanically driven systems are often employed for complex procedures, such as vascular, cardiac and neurosurgical procedures. Their movements, including rotations, panning and height adjustments, are motorized and tracked, enabling rapid repositioning during procedures. However, some C-arm intrinsic parameters, such as SID distance, may vary according to clinical requirements, thereby dynamically affecting the 3D navigation functionality. Therefore, for MT integration to be widely applicable in fluoroscopy-guided procedures, it is essential to accommodate both manually and mechanically driven C-arm systems. 
\end{enumerate}
In this study, we employ a tabletop radiolucent FG prototype~\cite{xia2023x} developed by Northern Digital Inc.. To the best of our knowledge, this is the first study to integrate a radiolucent FG into a fluoroscopy-guided workflow. To achieve this, we first design a two-layer radiolucent FG mounting frame. This hardware configuration is seamlessly integrated into the surgical table and ensures optimal MT accuracy. Subsequently, we propose a strategy of incorporating external fiducials onto the FG mounting frame to optimize the workflow and accommodate various C-arm systems. Specifically, external fiducials are employed to accurately estimate the C-arm pose and facilitate the registration of fluoroscopic images with pre-operative CTs, which further enables the development of 3D surgical navigation. In addition, via a video decomposition step, our framework significantly reduces imaging artifacts caused by metallic fiducials and the radio\-lucent FG, providing an unobstructed radiographic view under real-time fluoroscopy guidance.

%Stealth Station S8 with EM tracking~\footnote{\url{https://www.medtronic.com/ca-en/healthcare-professionals/products/neurological/surgical-navigation-systems/stealthstation/cranial-neurosurgery-navigation.html}}
% critical review of EM tracking in clinical setting~\cite{SPM+2020}, not needed if space is limited.

\section{Methods}\label{methods}
We begin by describing the hardware integration of the magnetic FG in Section~\ref{sec:hardware}. Then, MT-based 2D and 3D fluoroscopy navigations are described in Sections~\ref{sec:navigation} and~\ref{sec:3dnavigation}. Section~\ref{sec:integration} introduces the system integration. Finally, we introduce several metrics to evaluate the errors inherent to fluoro-CT registration. 

\subsection{Hardware Design}
\label{sec:hardware}
\begin{figure}[h!]
\centering
\begin{tabular}{@{} c @{} c @{}}
\includegraphics[width=.22\textwidth]{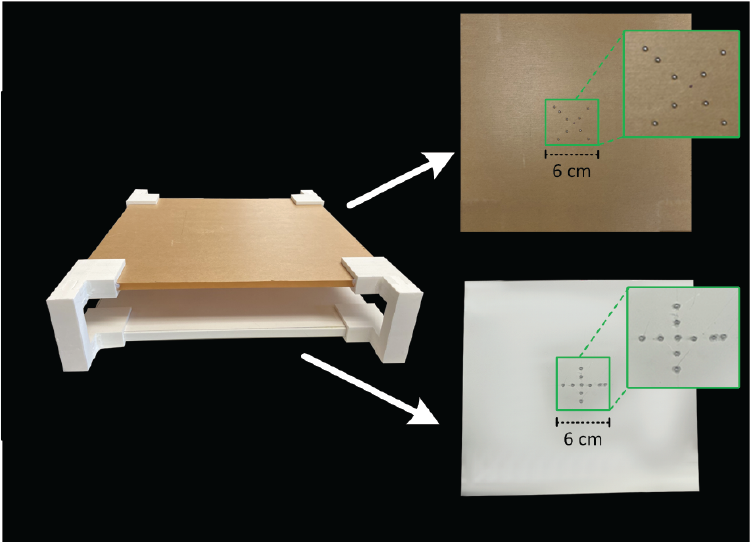}&
\includegraphics[width=.22\textwidth]{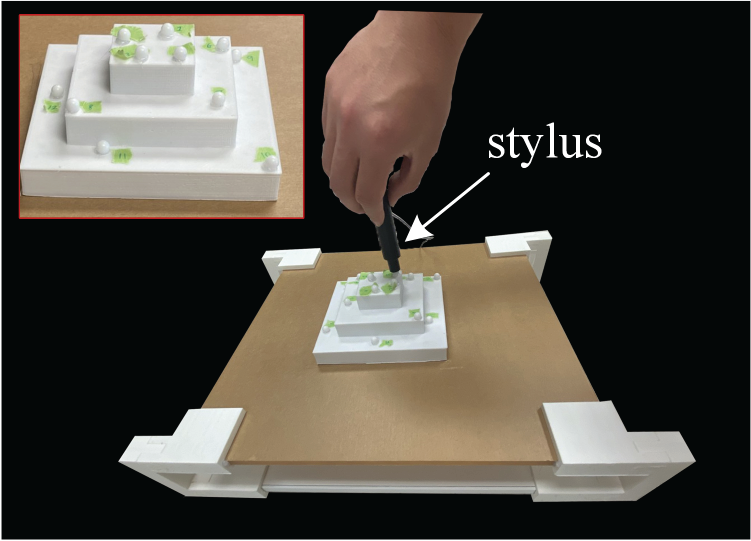}
\\
{(A)} & {(B)}\\

\end{tabular}
\caption{The design of the FG mounting frame (A) and the FG mounting frame with the pyramid phantom (B). 13 acetal resin spheres were attached to the pyramid phantom.}
\label{fig_NDI_FG_mounting_frame}
\end{figure}

The X-ray radiolucent FG prototype used in this study reduces metal-induced imaging artifacts compared to conventional MT systems. In addition, the prototype features a Metal Immunity Mode (MiM), which dramatically improves the resilience of the system tracking to metal distorters. Furthermore, the radiolucent FG prototype has a thinner profile than the previous window FG, thereby improving handling and system integration. The tabletop design also enables easy positioning on the surgical bed, as well as closer proximity to the patient's region of interest. Based on this prototype, we designed a two-layer FG mounting frame, comprising a square acrylic plate on the top and the FG on the bottom, as shown in Fig.~\ref{fig_NDI_FG_mounting_frame}A. This two-layer configuration was primarily designed to place the FG underneath the surgical bed, which maintained proximity to the surgical site with optimal tracking accuracy, as well as being non-intrusive within the surgical workspace. Our previous work\cite{xia2023x} detailed how external fiducial markers attached to the FG were utilized to estimate C-arm poses. In this regard, nineteen aluminum spherical fiducials with a diameter of \SI{4}{\milli\metre}, were attached to both the top (9 fiducials) and bottom (10 fiducials) layers within a \qtyproduct{6x6}{\centi\metre} region of interest. These external fiducials served as the foundation for robust pose estimation across different C-arm types and configurations. Fig.~\ref{fig_system_corodinate} illustrates the setup of the FG mounting frame in a clinical setting.

%\begin{figure}[h]
%\centerline{\includegraphics[width=18pc]{figs/NDI_FG_frame_v3.png}}
%\caption{The design of the FG mounting frame. The top layer comprises an acrylic plate, and the bottom layer houses the FG. Fiducials are attached to the downside of each layer.}
%\label{fig_NDI_FG_frame}
%\end{figure}

%\begin{figure}[h]
%   \centering
%    \begin{subfigure}{9.7pc}
%        \centering
        %\adjincludegraphics[width=\linewidth,trim={2cm 1cm 3cm 2cm},clip]{figs/mPD_LRAO.png}
%        \includegraphics[width=\linewidth] {figs/NDI_FG_frame_v3.png}
%        \caption{}
%    \end{subfigure}
%    \hfill
%    \begin{subfigure}{10pc}
%        \centering
        %\adjincludegraphics[width=\linewidth,trim={2cm 1cm 3cm 2cm},clip]{figs/mPD_CRAU.png}
%        \includegraphics[width=\linewidth] {figs/FG_frame_registration_CT2MT.png}
%        \caption{}
%    \end{subfigure}
%    \caption{The design of the FG mounting frame (a) and the FG mounting frame with the pyramid phantom (b). 13 acetal resin spheres were attached to the pyramid phantom.}
%    \label{fig_NDI_FG_mounting_frame}
%\end{figure}

\subsection{2D Navigation}
\label{sec:navigation}
In conventional 2D navigation, repeated fluoroscopic imaging to confirm instrument position leads to cumulative radiation exposure, posing safety risks for both the patient and radiologist. To overcome this limitation, we used MT for instrument tracking, minimizing the need for multiple fluoroscopic views. In the following sections, we briefly summarize our previous work for removing shadows of FG coils and aluminum fiducials in fluoroscopic images, followed by describing our virtual roadmap approach.

%\begin{figure}[h]
%    \centering
%    \begin{subfigure}{6.8pc}
%        \centering
%        \includegraphics[width=\linewidth]{figs/xray_1.png}
%        \caption{}
%    \end{subfigure}
%    \hfill
%    \begin{subfigure}{6.8pc}
%        \centering
%        \includegraphics[width=\linewidth]{figs/xray_2.png}
%        \caption{}
%    \end{subfigure}
%    \hfill
%    \begin{subfigure}{6.8pc}
%        \centering
%        \includegraphics[width=\linewidth]{figs/xray_3.png}
%        \caption{}
%    \end{subfigure}
    
%    \caption{An example of fluoroscopy decomposition. (a) X-ray image of a torso phantom and FG. The dark dots are aluminum fiducials attached to the FG; (b) decomposed residual image; (c) decomposed artifact-free image. }
%    \label{fig_xray_decomposition}
%\end{figure}

\begin{figure}[hptb] % changed from h!
\centering
\begin{tabular}{@{} c @{} c @{} c @{}}
\includegraphics[width=.16\textwidth]{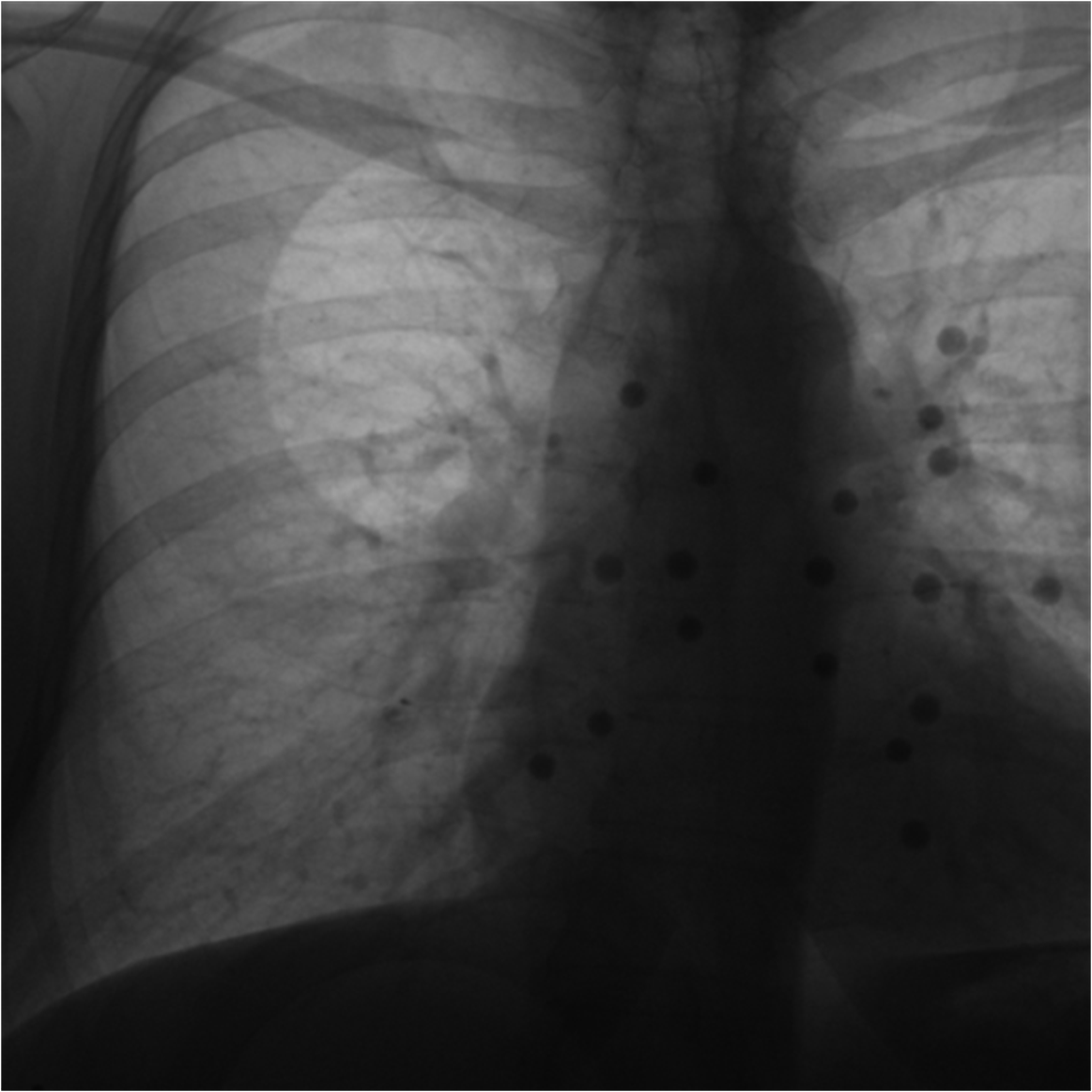}&
\includegraphics[width=.16\textwidth]{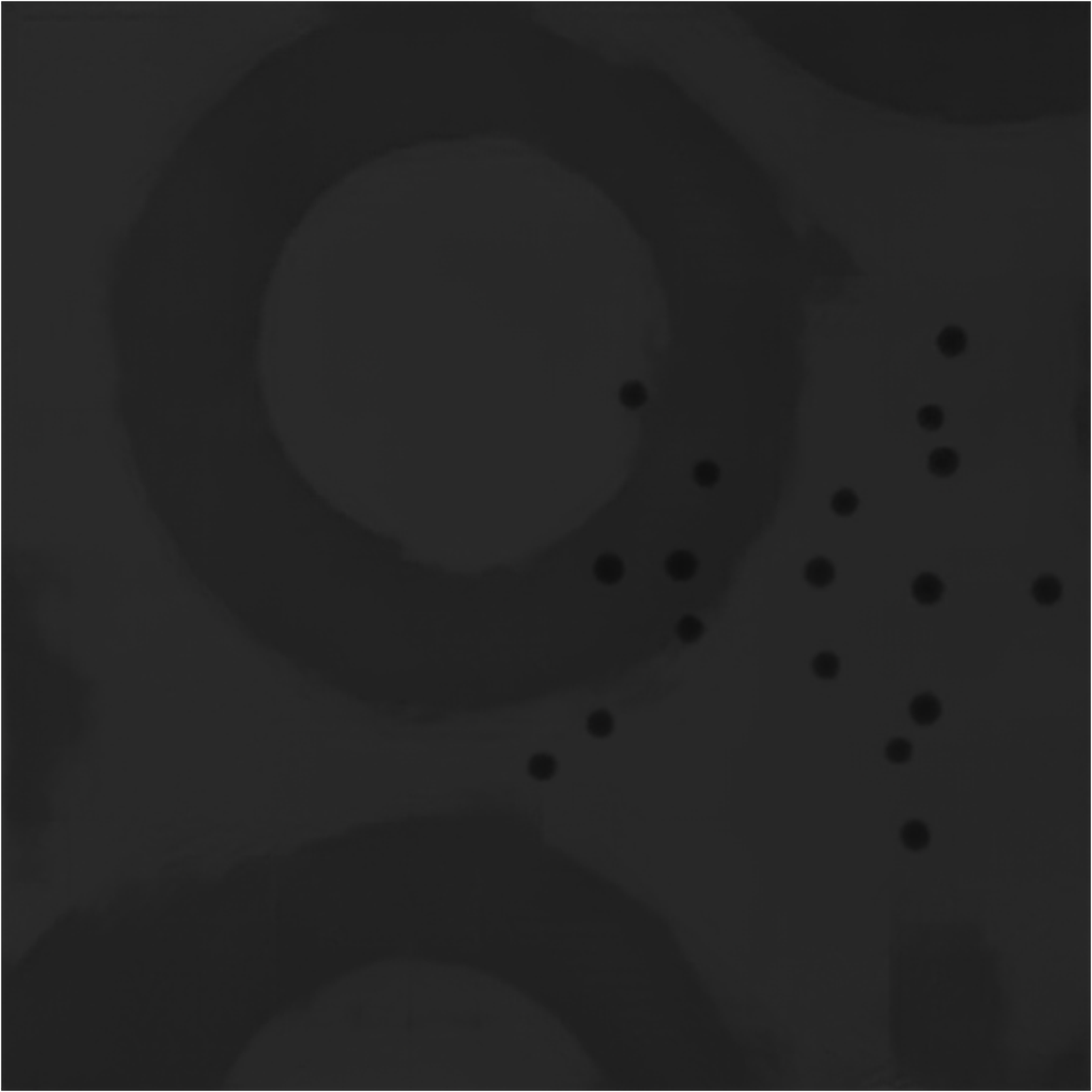}&
\includegraphics[width=.16\textwidth]{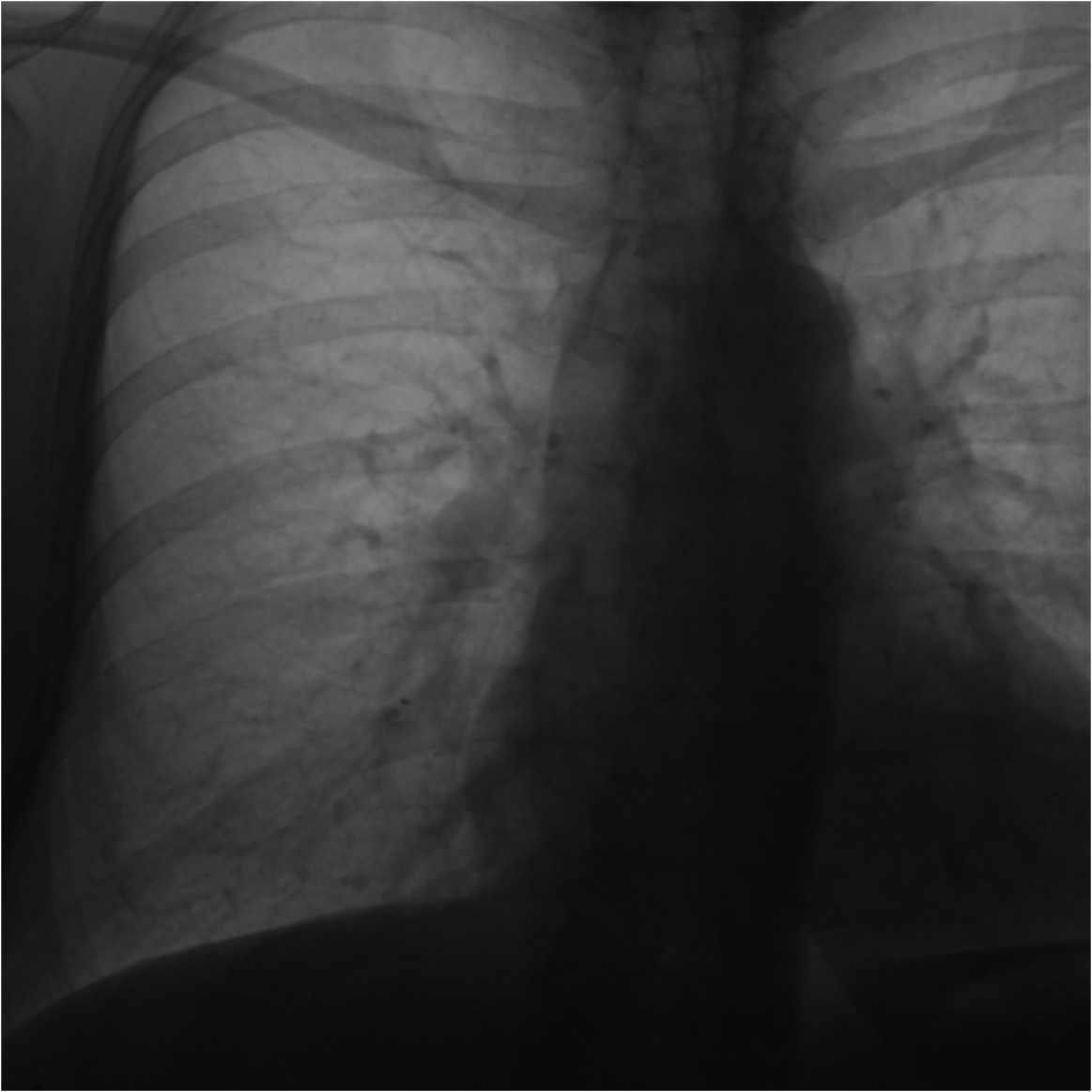}
\\
{(A)} & {(B)} & {(C)}
\\
\end{tabular}
\caption{An example of fluoroscopy decomposition. (A) X-ray image of a torso phantom and FG. The dark dots are aluminum fiducials attached to the FG; (B) decomposed residual image; (C) decomposed artifact-free image.}
\label{fig_xray_decomposition}
\end{figure}

\subsubsection{Fluoroscopy Decomposition}
We used a Deep Adversarial Decomposition approach\cite{xia2023x, zou2020deep} to eliminate image artifacts caused by the FG and radiopaque fiducials. The input to our network was the patient fluoroscopic image with the FG mounting frame, as shown in Fig.~\ref{fig_xray_decomposition}A. The two outputs comprised the residual image containing the extracted artifacts (Fig.~\ref{fig_xray_decomposition}B) and the artifact-free patient fluoroscopic image (Fig.~\ref{fig_xray_decomposition}C). Fiducials (dark dots induced by the aluminum spheres) were easily identified in the residual image due to their high contrast appearance. As the details of this fluoroscopy decomposition technique were discussed in our previous work~\cite{xia2023x}, this paper focuses on the navigation problem.

\begin{figure}[hptb] % changed from h!
\centering
%\centerline{\includegraphics[width=20pc]{figs/system_coordinates_v1.png}}
\includegraphics[width=20pc,trim={0cm 0cm 0cm 1cm},clip]{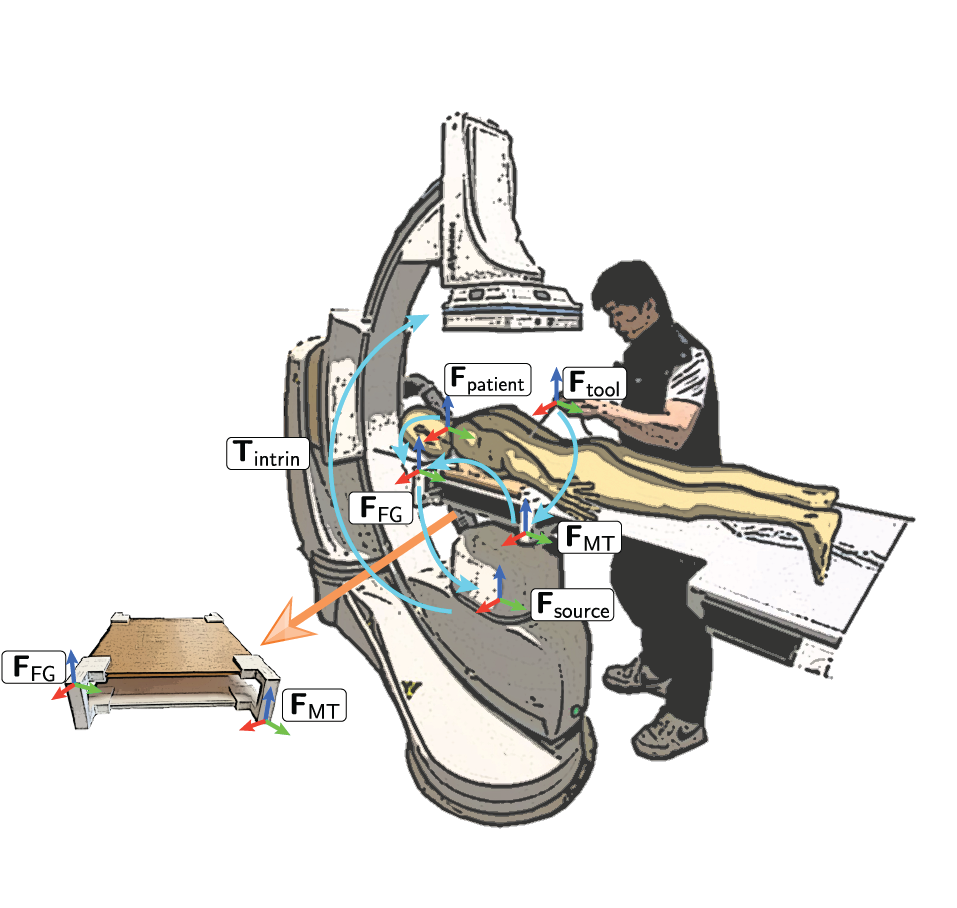}
%\centerline{\adjincludegraphics[width=\linewidth,trim={0cm 0cm 3cm 0cm},clip]{figs/system_coordinates_v1.png}}
\caption{System setup and coordinate frames for 2D and 3D navigation.}
\label{fig_system_corodinate}
\end{figure}

\subsubsection{Virtual Roadmap of Surgical Instruments}
The virtual instrument roadmap provides real-time pose information of the surgical instrument on static fluoroscopic images, while reducing continuous radiation exposure. To clarify the workflow, several coordinate frames were defined in Fig.~\ref{fig_system_corodinate}, including the physical FG mounting frame ($F_{FG}$), the MT coordinate frame ($F_{MT}$) to which the MT field is referenced, the patient coordinate frame ($F_{patient}$), the X-ray source frame ($F_{source}$) and the tracked instrument frame ($F_{tool}$). The mathematical model for the 2D navigation environment is expressed as:
\begin{equation}\label{eq1}
    [rot, t]_{2d} = T_{FG-3d}^{2d} * T_{MT}^{FG} * T_{tool}^{MT}
\end{equation} 
where $T_{tool}^{MT}$is a 4$\times$4 transformation matrix from the surgical instrument frame ($F_{tool}$) to $F_{MT}$, obtained from the tracking device. $T_{MT}^{FG}$ represents the registration transformation from $F_{MT}$ to $F_{FG}$. $T_{FG-3d}^{2d}$ is a 3$\times$4 projection matrix from the physical FG mounting frame to the 2D fluoroscopic image space. To achieve the virtual roadmap task, the matrices $T_{MT}^{FG}$ and $T_{FG-3d}^{2d}$ must be determined.

%\begin{figure}[h]
%\centerline{\includegraphics[width=18pc]{figs/FG_frame_registration_CT2MT.png}}
%\caption{The FG mounting frame with the pyramid phantom. 13 acetal resin spheres were attached to the pyramid phantom.}
%\label{fig_FG_frame_registration}
%\end{figure} 

$T_{MT}^{FG}$. We employed a point-to-point registration approach to align the physical FG mounting frame ($F_{FG}$) with the MT frame ($F_{MT}$). A three-layer pyramid embedded with 13 acetal resin spheres was designed and affixed on the FG mounting frame, as shown in Fig.~\ref{fig_NDI_FG_mounting_frame}B. Subsequently, an MT stylus with an inverted cone shape was used to capture the centroids of these spheres relative to $F_{MT}$. Simultaneously, sphere centroids relative to the CT image (or $F_{FG}$) were automatically extracted from the thresholded images. Finally, the rigid transformation, $T_{MT}^{FG}$, was calculated using singular value decomposition~\cite{klema1980singular}.

$T_{FG-3d}^{2d}$. To calculate the projection matrix, a critical step is to establish correspondence between the 2D fluoroscopic image and 3D CT image. Aluminum fiducials attached to the FG mounting frame, served as shared landmarks in both 2D and 3D spaces. In 2D space, the landmark centroids were automatically extracted from the residual image (see Fig.~\ref{fig_xray_decomposition}B) using the OpenCV “findContours()” function\footnote{https://opencv.org/}. In 3D, the landmark locations with respect to the FG mounting frame were identified from the CT image. These two sets of landmark measurements were related by a homogeneous transformation matrix to achieve the landmark matching. Finally, the Direct Linear Transformation algorithm~\cite{hartley2003multiple} was used to estimate the projection matrix $T_{FG-3d}^{2d}$.

\subsection{3D Navigation}
\label{sec:3dnavigation}
3D navigation aims to visualize the real-time spatial relationship between volumetric modalities and surgical instruments. In addition, 3D reformations, such as axial, sagittal, or coronal views, can be generated to facilitate the guidance procedure. A critical step in 3D navigation is determining the transformation $T_{tool}^{patient}$, which represents the instrument pose relative to the patient coordinate frame $F_{patient}$, as described in Eqn.~\eqref{eq2}. Given that the approach to compute $T_{MT}^{FG}$ is outlined in Section~\ref{sec:navigation}, the transformation $T_{patient}^{source}$ and $T_{FG}^{source}$ remain as the unknowns.
\begin{equation}\label{eq2}
    \begin{split}
    T_{tool}^{patient} &= T_{MT}^{patient}*T_{tool}^{MT} \\
    &= (T_{patient}^{source})^{-1} * T_{FG}^{source} * T_{MT}^{FG} * T_{tool}^{MT}
    \end{split}
\end{equation} 

\begin{figure*}[h]
\centerline{\includegraphics[width=35pc]{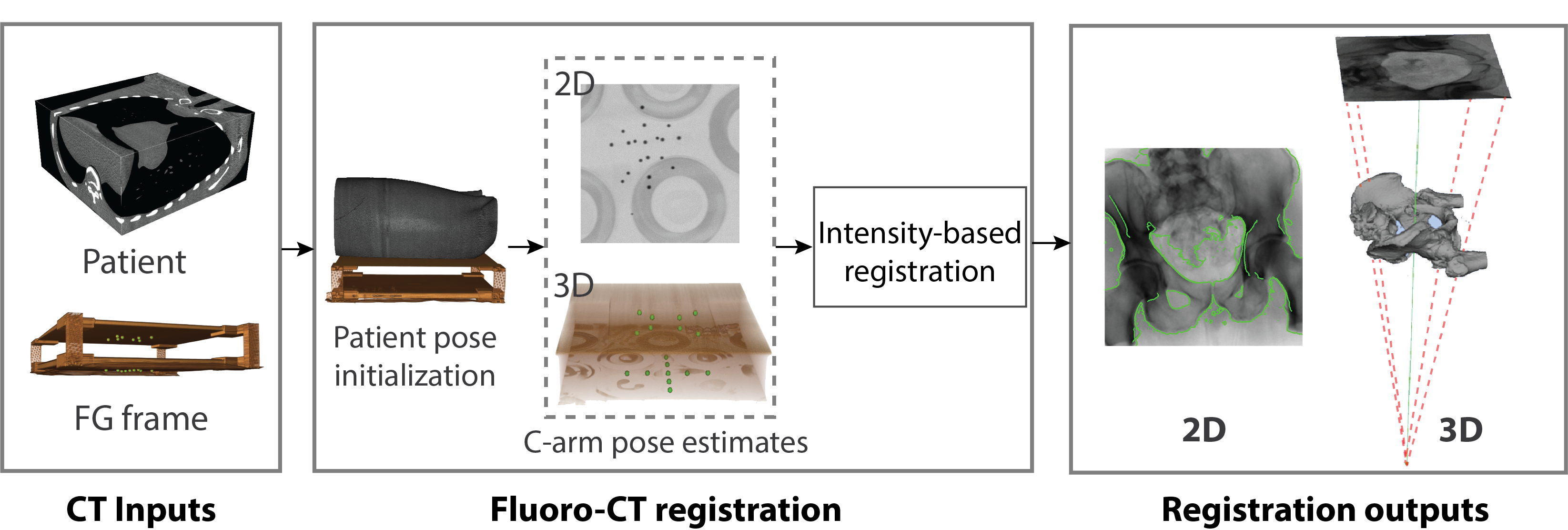}}
\caption{Workflow of fluoro-CT registration.}
\label{fig_workflow_fluoro_CT_registration}
\end{figure*}

$T_{patient}^{source}$. When combined with the known C-arm intrinsic parameters, this task becomes the classic fluoro-CT registration problem. Fig.~\ref{fig_workflow_fluoro_CT_registration} illustrates our proposed workflow to register fluoroscopic images with CT images, which encompasses three steps: patient pose initialization, C-arm pose estimation (\ie, $T_{FG}^{source}$), and intensity-based registration. 
\begin{enumerate}
    \item \textit{Patient pose initialization}. Fluoroscopy-guided procedures target various anatomical regions. This step involves positioning the patient’s region of interest on the FG mounting frame according to clinical requirements, as shown in Fig.~\ref{fig_workflow_fluoro_CT_registration}. After initializing the patient pose, the relative relationship between $F_{patient}$ and $F_{FG}$ in the CT image space should approximate the actual clinical setup. This step has a high degree of misalignment tolerance and is only executed once. As long as the patient CT is approximately aligned, accurate fluoro-CT registration can be achieved using the subsequent steps detailed below.
    \item \textit{C-arm pose estimation} ($T_{FG}^{source}$). The C-arm pose refers to the pose of the FG mounting frame relative to the C-arm source frame $F_{source}$. The C-arm intrinsics, denoted as $T_{intrin}$, were derived from C-arm parameters, including pixel spacing, image size, and SID distance (\ie, focal length). The C-arm pose was determined by decomposing the known projection matrix ($T_{FG-3d}^{2d}=T_{intrin}*T_{FG}^{source}$). To estimate the C-arm pose, we used a Perspective-n-Point (PnP) approach~\cite{hartley2003multiple}. Note that Section~\ref{sec:integration} described how C-arm parameters are automatically obtained.
    \item \textit{Intensity-based registration}. Beginning with the initial C-arm pose, an image intensity-based multi-resolution registration strategy~\cite{grupp2020automatic, grupp2019pose} was employed to optimize the patient pose parameters (\ie, $T_{patient}^{source}$). This was accomplished by using similarity metrics to compare simulated fluoroscopic images, derived from digitally reconstructed radiographs (DRRs), with the real fluoroscopic image $I_{fluoro}$, as given in Eqn.~\eqref{eq3}. 
    DRRs ($\mathcal{D}$) were generated by ray casting through the CT image ($I_{ct}$), based on the intrinsic $T_{intrin}$ and the current extrinsic $\tilde{T}_{patient}^{source}$ parameters. Due to variations in peak kilovolt (kVp) settings between $I_{ct}$ and $I_{fluoro}$, a similarity metric $\mathcal{L}$, (patch-based gradient normalized cross correlation~\cite{grupp2020automatic}), was employed as the objective function to guide the optimization process. To determine the optimal patient pose parameters while minimizing computational workload, we implemented a two-step registration framework. Initially, a coarse registration step was performed using images that were down-sampled by a factor of 8 in each 2D dimension. A derivative-free optimizer, “Covariance Matrix Adaptation: Evolutionary Search” (CMA-ES)~\cite{hansen2001completely}, was applied due to its robustness to local minima. This was followed by a fine-level registration with images down-sampled by a factor of 4, using another derivate-free optimizer, “Bounded Optimization by Quadratic Approximation (BOBYQA)”~\cite{powell2009bobyqa} to expedite convergence. The C-arm pose $T_{FG}^{source}$ was also used to initialize the transformation $\tilde{T}_{patient}^{source}$.
    
\end{enumerate}

\begin{equation}\label{eq3}
     T_{patient}^{source} = \arg\min_{\tilde{T}_{patient}^{source}} \mathcal{L}[{\mathcal{D}(I_{ct}, T_{intrin}, \tilde{T}_{patient}^{source}), I_{fluoro}}]
\end{equation} 
    
\begin{figure*}[h]
\centerline{\includegraphics[width=35pc]{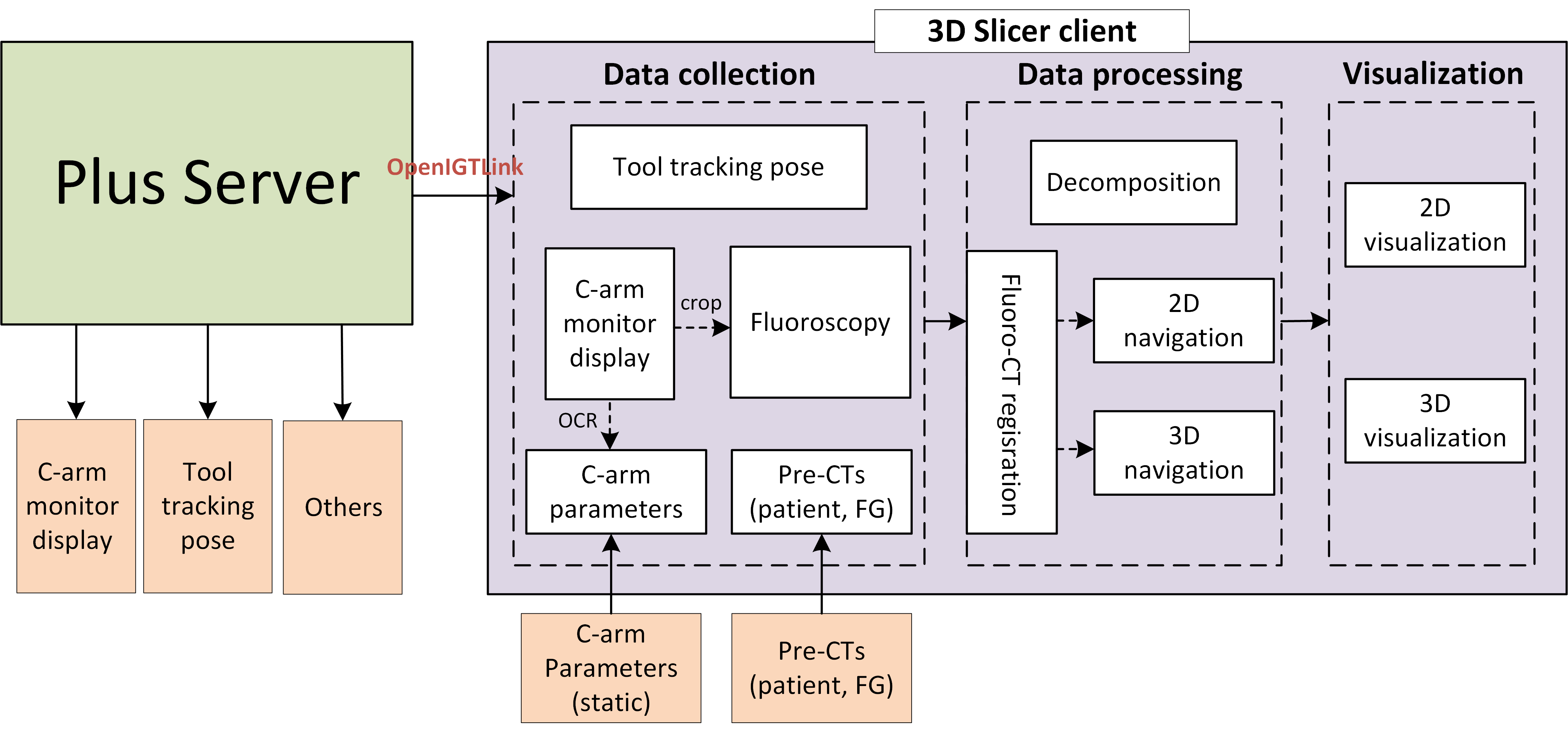}}
\caption{Our server-client architecture for fluoroscopy-guided procedures.}
\label{fig_workflow_system_integration}
\end{figure*}

\subsection{System Integration}
\label{sec:integration}
Typically, C-arm parameters and fluoroscopic images can be transferred to a local DICOM server and retrieved by a client for downstream tasks, such as fluoro-CT registration. However, this approach often fails to meet real-time performance requirements. Moreover, integrating all the multi-modal inputs, such as the real-time instrument tracking data, diagnostic CTs and fluoroscopic images, into a single platform is essential for efficient fluoroscopy-guided workflows. To address these challenges, we developed a 3D Slicer\footnote{https://www.slicer.org/}-based server-client architecture that seamlessly streams multi-modal inputs and is compatible with various C-arm systems (Fig.~\ref{fig_workflow_system_integration}).

First, we used PlusServer\footnote{https://plustoolkit.github.io/}, a standalone server application, to acquire data from multiple hardware devices, including C-arms and spatial trackers. For C-arm inputs, a frame grabber (Epiphan Video DVI2USB 3.0, Ottawa, Canada) captured the content displayed on the C-arm monitor at 30 frames per second (fps). In addition to the fluoroscopic image, C-arm parameters (\eg, the SID) that are shown within the clinical display for mechanically driven C-arm systems, were also captured in this video stream input. For instrument pose tracking input, PlusServer was configured to acquire multi-channel tracking data at a streamming rate of 30 fps. Additional inputs, such as the diagnostic CTs, the FG mounting frame CT and default C-arm parameters, were sent directly to the client (3D Slicer) to reduce data streaming workload. 

Then, PlusServer streamed the multi-modal data to our 3D Slicer client via OpenIGTLink\footnote{https://openigtlink.org/}, a lightweight, TCP/IP-based communication protocol specifically developed for live data transfer in image-guided interventions. 3D Slicer’s built-in OpenIGTLink module enabled data retrieval from PlusServer, while the 3D Slicer platform supports visualization, image processing, and data analysis, making it an ideal software platform for our downstream tasks.

And finally, our 3D Slicer client module comprised three components: data collection, processing and visualization, as detailed in Fig.~\ref{fig_workflow_system_integration}. In data collection, the fluoroscopy image was cropped from the captured monitor display, and C-arm parameters were automatically recognized using an optical character recognition (OCR) tool, Python-tesseract\footnote{https://pypi.org/project/pytesseract/}. This step was applicable only for mechanically driven C-arms, where C-arm parameters, such as SID, are subject to change during the intervention. Data processing follows the methods detailed in Sections~\ref{sec:navigation} and~\ref{sec:3dnavigation}, and visualization was demonstrated in the results section. 

\subsection{Evaluation Metrics}
Inspired by the work of Kraats~\etal~\cite{van2005standardized}, three metrics were used to evaluate registration accuracy: mean target registration error (mTRE), mean projection distance (mPD), and mean reprojection distance (mRPD). The target points in the patient CT space are denoted as $P_{i} (i=1,2,…,n)$. The ground-truth and calculated extrinsics are represented by $\dot{T}_{patient}^{source}$ and $T_{patient}^{source}$, respectively. The 3D registration error, mTRE, is given by:
\begin{equation}\label{eq4}
mTRE = \frac{1}{N} \sum_{i=1}^{N} \| \dot{T}_{patient}^{source} * P_{i} - T_{patient}^{source} * P_{i} \|
\end{equation} 
In fluoroscopic guidance, fluoro-CT registration is often employed to overlay 3D information, such as surgical plans, onto 2D fluoroscopic images. To evaluate the accuracy of this mapping, 2D registration metrics, including mPD and mRPD, are used. The mPD measures the distance between the 2D projections of target points at the registration position and the projections of target points at the ground-truth position, which can be defined as: 
\begin{equation}\label{eq5}
mPD = \frac{1}{N} \sum_{i=1}^{N} \| T_{intrin}*\dot{T}_{patient}^{source} * P_{i} - T_{intrin}*T_{patient}^{source} * P_{i} \|
\end{equation} 
The second 2D metric (mRPD) computes the minimum distance from the 3D target point (\ie, $\dot{T}_{patient}^{source} *P_{i}$) to the line that is defined from the X-ray source to the corresponding 2D projected point (\ie, $T_{intrin}*T_{patient}^{source}*P_{i}$). 

\section{Experiments and Results}
\subsection{Hardware effects on MT}
\subsubsection{Aluminum fiducials} Aluminum, a paramagnetic material, is weakly attracted to magnetic fields. To assess its effects on MT, we conducted a controlled experiment, whereby FG was placed in a non-ferromagnetic environment. In the experimental group, aluminum fiducials were attached to the FG, as shown in Fig.~\ref{fig_NDI_FG_mounting_frame}. Then a magnetically trackable needle, manufactured by NDI, was used to record 21 different tracking positions for both control and experimental groups. The positional distances between the two groups were calculated, with a mean tracking distance of \SI{0.02}{\milli\metre} $\pm$ \SI{0.01}{\milli\metre}. In addition, a statistical analysis using one-way ANOVA yielded a p-value of 0.53 $>$ 0.05, indicating no statistically significant difference between the groups.
\subsubsection{C-arm} We also investigated the effect of the C-arm on the accuracy of the MT system. In this experiment, the C-arm and the FG mounting frame attached with aluminum fiducials were introduced into the experiment. We recorded needle positions at 6 different SIDs, ranging from \SI{120}{\centi\metre} to \SI{95}{\centi\metre} in \SI{5}{\centi\metre} increments. Since the FG mounting frame was placed between the X-ray source and the image detector, these SIDs corresponded to distances of \SI{65}{\centi\metre} to \SI{40}{\centi\metre} from the FG to the image detector. For each distance, 15 positions were recorded with the C-arm powered on. The ground truth was defined as the configuration with a SID of \SI{120}{\centi\metre} and without the inclusion of aluminum fiducials. We measured the distance between the ground truth and the groups with different SIDs, as presented in Table~\ref{tab_effects_C-arm}. The group with a \SI{95}{\centi\metre} SID exhibited the largest distance error, with a mean of \SI{0.19}{\milli\metre} $\pm$ \SI{0.07}{\milli\metre}. Then, a multiple comparison Dunnett’s test~\cite{dunnett1955multiple}, was performed to analyze statistical differences between the control group (SID $=$ \SI{120}{\centi\metre}) and the other groups. The results indicated that groups with SIDs ranging from \SI{115}{\centi\metre} to \SI{90}{\centi\metre}, had p-values greater than 0.05 implying no statistically significant differences. However, the group with a \SI{95}{\centi\metre} SID had a p-value of 0.003, implying that that the accuracy of the MT is affected when the FG-to-detector distance is \SI{40}{\centi\metre}. Since the FG-to-detector distance in the actual clinical setting is generally larger than \SI{40}{\centi\metre}, this statistically significant difference would not have a clinically significant impact on interventions.

\begin{table*}[h!]
\caption{Results of effects of the C-arm on tracking accuracy.}
\centering
\setlength{\tabcolsep}{8pt}
\begin{tabular}{ccccccc}
\Xhline{2\arrayrulewidth}
{SIDs (cm)} & 120 & 115 & 110 & 105 & 100 & 95 \\
\hline
{Distance error (mm)} & 0.02 $\pm$ 0.03 & 0.02 $\pm$ 0.04 & 0.04 $\pm$ 0.05 & 0.06 $\pm$ 0.04 & 0.09 $\pm$ 0.03 & 0.19 $\pm$ 0.07 \\
$p$-value & {-} & $>$ 0.05 & $>$ 0.05 & $>$ 0.05 & $>$ 0.05 & $<$ 0.05 \\
\Xhline{2\arrayrulewidth}
\end{tabular}
\label{tab_effects_C-arm}
\end{table*}

\begin{table*}[h!]
\caption{Virtual augmentation error of tracked tool tips in fluoroscopy images.}
\centering
\setlength{\tabcolsep}{8pt}
\begin{tabular}{cccc}
\Xhline{2\arrayrulewidth}
{Metrics} & AP & (LAO 10 $^{\circ}$, CRA 1$^{\circ}$) & (RAO 10 $^{\circ}$, CAU 1$^{\circ}$) \\
\hline
{$\#$ of images} & 39 & 31 & 40 \\
{Tip Euclidean distance(mm)} & 0.60 $\pm$ 0.32 & 0.86 $\pm$ 0.44 & 0.77 $\pm$ 0.61 \\
{Signed tip distance(X)(mm)} & 0.35 $\pm$ 0.21 & 0.23 $\pm$ 0.14 & -0.33 $\pm$ 0.38 \\
{Signed tip distance(Y)(mm)} & -0.20 $\pm$ 0.51 & -0.76 $\pm$ 0.54 & -0.17 $\pm$ 0.84 \\
{Unsigned angle error($^{\circ}$)} & 0.67 $\pm$ 0.31 & 1.01 $\pm$ 0.19 & 0.36 $\pm$ 0.19\\
\Xhline{2\arrayrulewidth}
\end{tabular}
\label{tab_virtual_augmentation_error}
\end{table*}

\begin{table*}[h!]
\caption{Registration results across various clinical settings. Note that “phantom movement” simulates the patient’s movement during the procedure. (mPD: mean projection distance; SR: success rate) }
\centering
\setlength{\tabcolsep}{8pt}
\begin{tabular}{c |c |c c| c c| c c}
\Xhline{2\arrayrulewidth}
\multirow{2}{*}{C-arm settings} & \multirow{2}{*}{\# of images} & \multicolumn{2}{c|}{Na\"ive}&\multicolumn{2}{c|}{Pose-only} & \multicolumn{2}{c}{Ours} \\ 
{}& {} & mPD (mm) & SR & mPD (mm) & SR & mPD (mm) & SR\\
\hline

{LAO-RAO (-30$^{\circ}$ - 30$^{\circ}$)} & {54} & {0.55 $\pm$ 0.06} & {62.96\%} & {1.81 $\pm$ 0.07} & {94.44\%} & {0.56 $\pm$ 0.05} & {100\%} \\

{CRA-CAU (-30$^{\circ}$ - 30$^{\circ}$)} & {64} & {0.53 $\pm$ 0.03} & {59.38\%} & {1.84 $\pm$ 0.08} & {98.44\%} & {0.52 $\pm$ 0.05} & {100\%} \\

{SID (90 cm - 125 cm)} & {38} & {0.60 $\pm$ 0.04} & {84.21\%} & {0.82 $\pm$ 0.09} & {100\%} & {0.59 $\pm$ 0.04} & {100\%} \\

{Phantom movement} & {31} & {0.61 $\pm$ 0.04} & {61.29\%} & {-} & {-} & {0.61 $\pm$ 0.06} & {83.87\%} \\

\Xhline{2\arrayrulewidth}
\end{tabular}
\label{tab_registration_error}
\end{table*}

\subsection{2D Navigation: virtual roadmap of surgical instruments}
We evaluated the accuracy of the virtual surgical instrument roadmap. Three fluoroscopic videos with a tracked needle from different views, including anterior-posterior (AP), (left-anterior-oblique (LAO)~\SI{10}{\degree} $+$ cranial (CRA)~\SI{1}{\degree}), (right-anterior-oblique (RAO)~\SI{10}{\degree} $+$ caudal (CAU)~\SI{1}{\degree}), were acquired as the ground truth dataset. Following Section~\ref{sec:navigation}, the tracked needle was virtually projected onto the fluoroscopic videos. Table~\ref{tab_virtual_augmentation_error} shows the needle tip and angle errors between the tracked needle and the ground truth. The ground truth was established by manually annotating two points on the high-contrast needle in each fluoroscopic image: the needle tip $p_{tip}$, and an arbitrary point on the needle shaft $p_{shaft}$. The needle tip error is defined as the Euclidean and signed distances between the ground truth ($p_{tip}$) and the tracked needle tip. The needle angular error was calculated as the angle difference between the ground truth direction (\ie, $\overrightarrow{p_{shaft}*p_{tip}}$) and the tracked needle direction. As shown in Table~\ref{tab_virtual_augmentation_error}, the mean Euclidean distance of the needle tip is approximately \SI{0.7}{\milli\metre}, and the mean needle angular error is less than \SI{1}{\degree}. In addition, the signed distance of the needle tip indicates a view-dependent bias along the X and Y axes of the fluoroscopic image.  
\subsection{3D Navigation: fluoro-CT registration}
The accuracy and robustness of the fluoro-CT registration were evaluated using a phantom designed for training in “endoleak” repair procedures~\cite{chen2015management}. The registration was performed across various clinical settings, including different angular and orbital rotation angles, SIDs, and simulated patient movements. To further assess generalizability, the method was tested on two public clinical datasets: cerebral angiograms\footnote{https://lit.fe.uni-lj.si/en/research/resources/3D-2D-GS-CA/} and pelvic images\footnote{https://github.com/rg2/DeepFluoroLabeling-IPCAI2020}. 
\subsubsection{Evaluation on endoleak phantom}
The endoleak phantom contains spinal elements, an aortic stent, a simulated aneurysm and feeding vessels, embedded in polyvinyl alcohol cryogel (PVA-C) as a tissue mimic. In the endoleak repair procedure, a needle-catheter combination is guided into the feeding and draining vessels to occlude the persistent perfusion. Fluoroscopic images were acquired under various settings, including angular rotation (\SI{-30}{\degree} to \SI{30}{\degree}), orbital rotation (\SI{-30}{\degree} to \SI{30}{\degree}), SIDs (\SI{90}{\centi\metre} to \SI{125}{\centi\metre}) and phantom motion, to evaluate the accuracy, robustness and capture range of our registration approach. The 2D X-ray images were captured with an image size of 1017 $\times$ 1017 pixels and a pixel spacing of \qtyproduct{0.29 x 0.29}{\milli\metre}. Target points (20-23) were selected from the stent intersection points (see Fig.~\ref{fig_fluoro_CT_registration_datasets}A). The mPD was used to measure the 2D guidance error, while the success rate (SR) was reported as the percentage of mPD measurements less than 5 mm.

The performance of fluoro-CT registration was evaluated by comparing our approach with two baseline approaches. \begin{inparaenum}
\item The na\"ive approach, which initializes the C-arm pose using the default AP view, followed by image intensity-based registration step to align the fluoroscopic images with the CT image; and
\item The pose-only approach, where the C-arm pose, $T_{FG}^{source}$, was estimated using PnP approach based on 2D/3D landmark correspondence. A correction transformation ($T_{patient}^{FG}$) was required to complete the fluoro-CT registration, calculated using $T_{patient}^{FG}=(T_{FG}^{source})^{-1}*T_{patient}^{source}$. (Note that this step only needs to be performed once)
\end{inparaenum}

Table~\ref{tab_registration_error} shows registration results across various clinical settings. Our approach achieved mPDs within the range of \SI{0.5}{\milli\metre} to \SI{0.6}{\milli\metre}, which is comparable to the na\"ive approach but with significantly higher SRs, achieving \SI{100}{\percent} success in all cases except one (\SI{83.9}{\percent} vs. \SI{61.3}{\percent}). Although the pose-only approach yielded slightly lower SRs than ours, its mPDs (\SI{1.81}{\milli\metre} $\pm$ \SI{0.07}{\milli\metre} and \SI{1.84}{\milli\metre} $\pm$ \SI{0.08}{\milli\metre}) were more than three times higher than ours in the LAO-RAO and CRA-CAU settings. Additionally, the pose-only approach does not account for patient movement during the procedure.

%\begin{figure}[h]
%    \centering
%    \begin{subfigure}{10pc}
%        \centering
%        \adjincludegraphics[width=\linewidth,trim={2cm 1cm 3cm 2cm},clip]{figs/mPD_LRAO.png}
        %\includegraphics[width=\linewidth]{figs/mPD_LRAO.png}
%        \caption{}
%    \end{subfigure}
%    \hfill
%    \begin{subfigure}{10pc}
%        \centering
%       \adjincludegraphics[width=\linewidth,trim={2cm 1cm 3cm 2cm},clip]{figs/mPD_CRAU.png}
        %\includegraphics[width=\linewidth]{figs/mPD_CRAU.png}
%        \caption{}
%    \end{subfigure}
%    \caption{Registration results of different approaches along angular (a) and orbital directions (b). }
%    \label{fig_mPD_angular_orbital}
%\end{figure}

\begin{figure}[h!]
\centering
\begin{tabular}{@{} c @{} c @{}}
\adjincludegraphics[width=.23\textwidth,trim={2cm 1cm 3cm 2cm},clip]{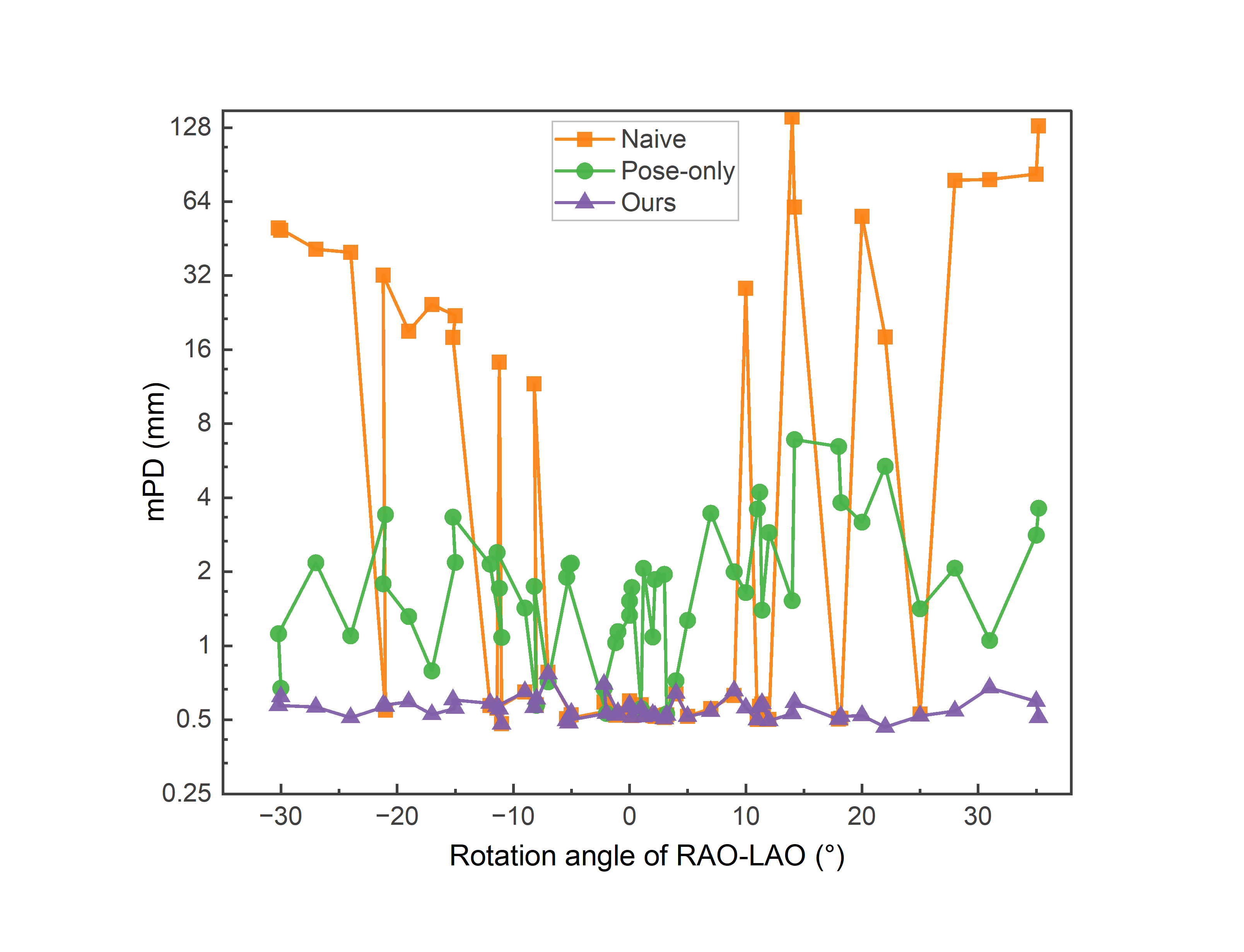}&
\adjincludegraphics[width=.23\textwidth,trim={2cm 1cm 3cm 2cm},clip]{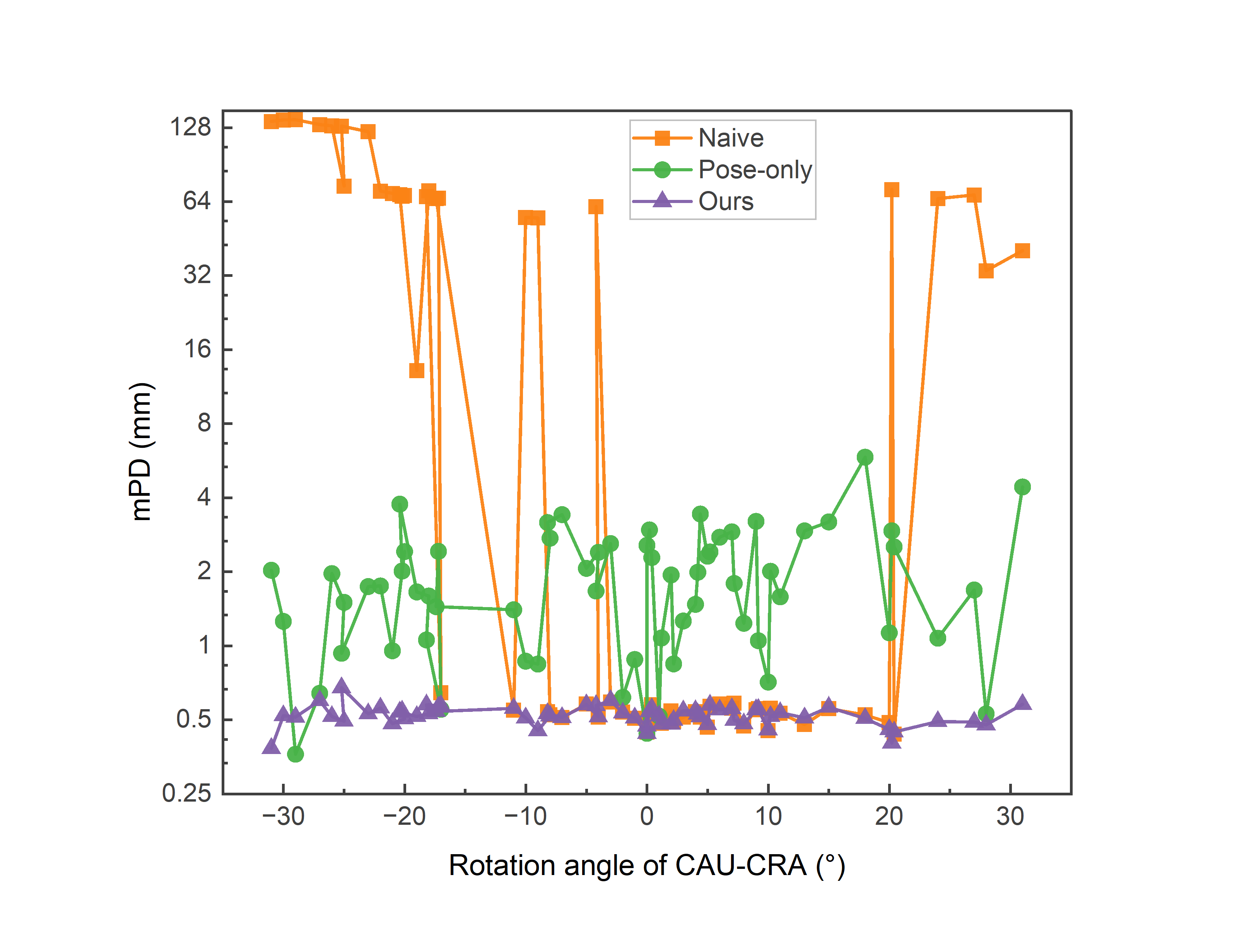}
\\
{(A)} & {(B)}\\
\end{tabular}
\caption{Registration results of different approaches along angular (A) and orbital directions (B).}
\label{fig_mPD_angular_orbital}
\end{figure}

Fig.~\ref{fig_mPD_angular_orbital} shows the capture range of different registration approaches with respect to angular and orbital rotations. In Fig.~\ref{fig_mPD_angular_orbital}A, the na\"ive approach initialized at approximately \SI{0}{\degree}, achieved successful alignment ($<$ \SI{5}{\milli\metre}) only within a limited range of \SI{-10}{\degree} to \SI{10}{\degree}. The pose-only method demonstrated a broader capture range (\SI{-30}{\degree} to \SI{30}{\degree}), but exhibited higher registration errors, with an mPD of approximately \SI{2}{\milli\metre}. In contrast, our approach consistently achieved a low mPD of approximately \SI{0.5}{\milli\metre}, independent of angular rotation angles. Similar results were observed for rotations along the CRA-CAU axis, as shown in Fig.~\ref{fig_mPD_angular_orbital}B.

%\begin{figure}[h]
%    \centering
%    \begin{subfigure}{10pc}
%        \centering
%        \adjincludegraphics[width=\linewidth,trim={2cm 1cm 3cm 2cm},clip]{figs/mPD_phantom_movements.png}
        %\includegraphics[width=\linewidth]{figs/mPD_LRAO.png}
%        \caption{}
%    \end{subfigure}
%    \hfill
%    \begin{subfigure}{10pc}
%        \centering
%        \adjincludegraphics[width=\linewidth,trim={2cm 1cm 3cm 2cm},clip]{figs/mPD_number_of_fiducials.png}
        %\includegraphics[width=\linewidth]{figs/mPD_CRAU.png}
%        \caption{}
%    \end{subfigure}
%    \caption{Registration results of different patient/phantom movements (a) and different number of fiducials (b). }
%    \label{fig_mPD_movement_fiducials}
%\end{figure}

\begin{figure}[h!]
\centering
\begin{tabular}{@{} c @{} c @{}}
\adjincludegraphics[width=.23\textwidth,trim={2cm 1cm 0.3cm 1cm},clip]{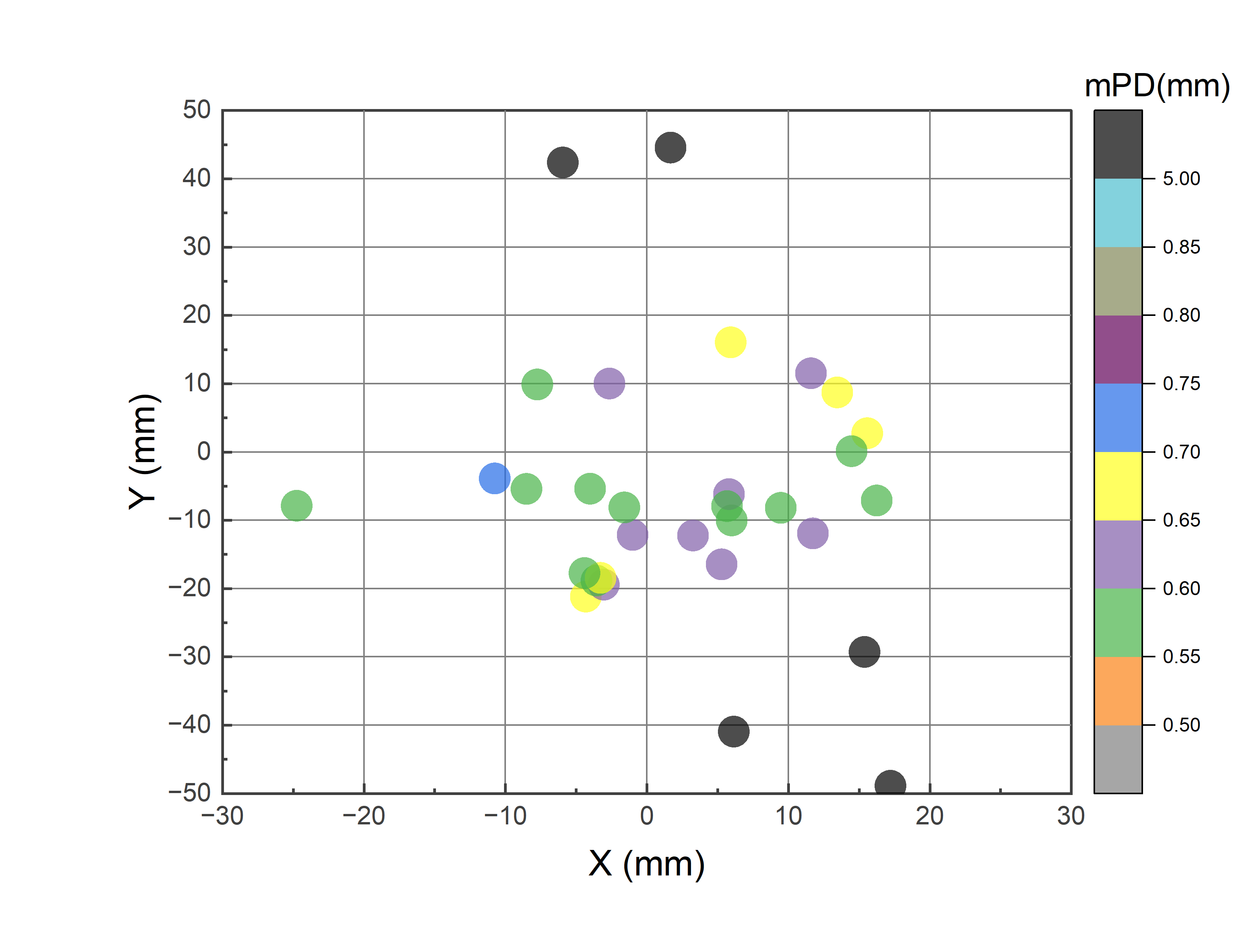}&
\adjincludegraphics[width=.23\textwidth,trim={2cm 1cm 0.3cm 1cm},clip]{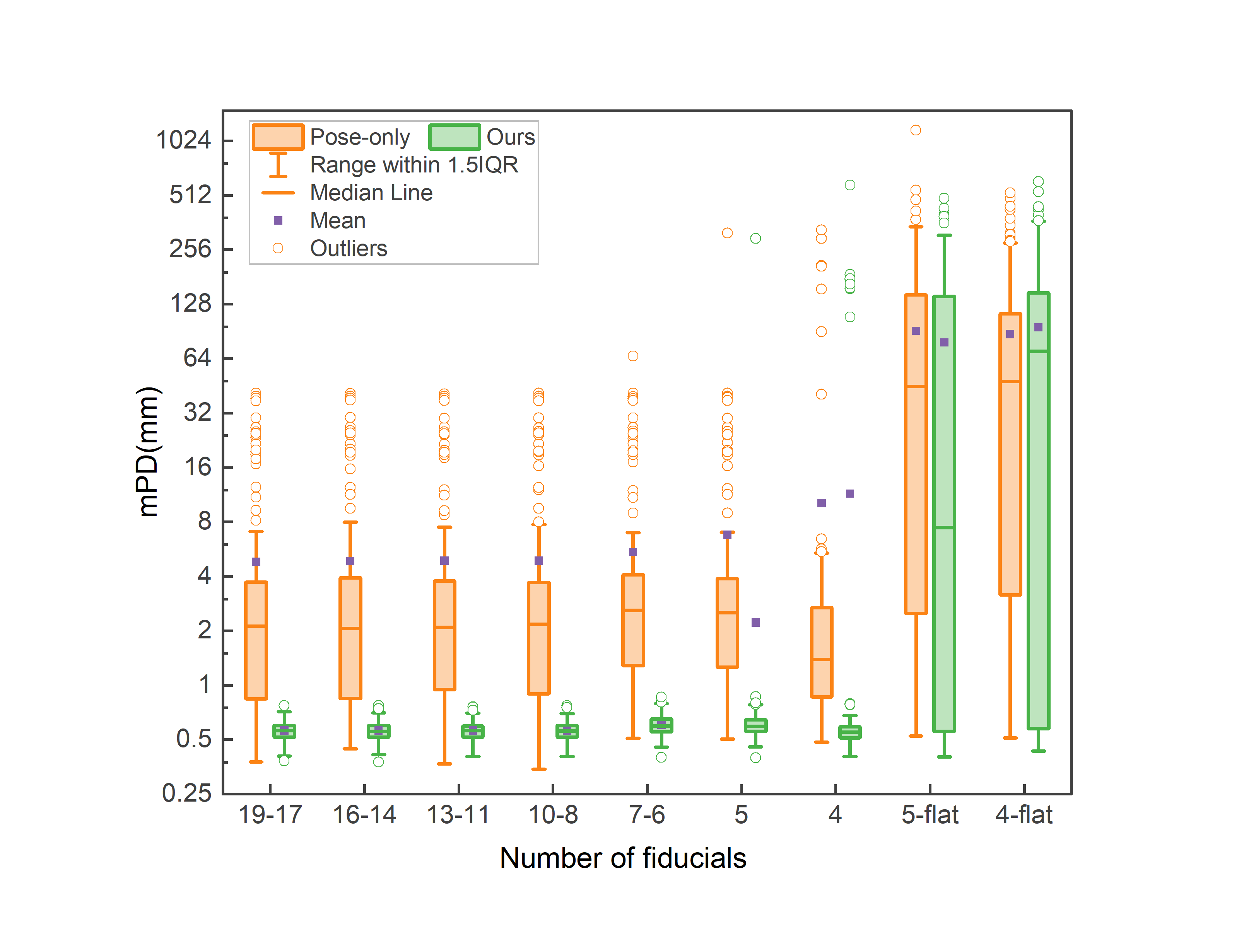}
\\
{(A)} & {(B)}\\
\end{tabular}
\caption{Registration results of different patient/phantom movements (A) and different number of fiducials (B). }
\label{fig_mPD_movement_fiducials}
\end{figure}

Patient movement during the procedure, due to discomfort or breathing, can cause significant misalignment in registered fluoroscopic images, potentially leading to erroneous instrument guidance. To evaluate the robustness of our registration approach under different amounts of movement, we conducted an experiment using the endoleak phantom, initially centered at (x = 0, y = 0), which was randomly translated and rotated within a range of \qtyproduct{30 x 50}{\centi\metre} and \qtyrange{0}{5}{\degree}. As shown in Fig.~\ref{fig_mPD_movement_fiducials}A, 26 out of 31 cases were successfully registered, all of which fell within a \qtyproduct{30 x 30}{\milli\metre} region. Beyond this range, the validity of our approach may be compromised.

Given the projective nature of X-ray fluoroscopy, fiducials may be obstructed or fall outside of the imaging field of view during the intervention. To assess the impact of the number of imaged fiducials on registration accuracy, we used 156 fluoro-CT image pairs acquired under various C-arm settings, including LAO-RAO (54 pairs), CRA-CAU (64 pairs), and different SIDs (38 pairs). The number of fiducials used for C-arm pose estimation was divided into 9 groups: 19-17, 16-14, 13-11, 10-8, 7-6, 5, 4, 5-flat and 4-flat. The “5-flat” and “4-flat” groups refers to fiducials located on a single layer (either top or bottom), while the other groups consisted of fiducials randomly selected from both layers. As shown in Fig.~\ref{fig_mPD_movement_fiducials}B, both our approach and the pose-only approach demonstrated consistent and accurate mPDs with $\geq$ 6 fiducials. Although good alignment was still possible with 4 or 5 fiducials, robustness decreased, as evidenced by outliers. Neither approach achieved acceptable registration accuracy with the “4-flat” or “5-flat” configurations, indicating that a multi-layered fiducial configuration is suggested to achieve robust registration. 
\subsubsection{Evaluation on public datasets}
Other than our local endoleak dataset, two public clinical datasets (\ie, \textit{3D-2D-GS-CA} and \textit{DeepFluoro}) were used to evaluate the generalization of our fluoro-CT registration workflow.

\textit{3D-2D-GS-CA}: This dataset includes images from 10 patients who underwent cerebral-endovascular procedures, with 8 cases focused on aneurysm treatment and 2 on arteriovenous malformation. All images were acquired intraoperatively just prior to treatment. For each patient, a 3D digitally subtracted angiogram (3D-DSA) with vessel contrast was acquired, which had voxel sizes of \qtyproduct{0.46 x 0.46 x 0.46}{\milli\metre} and image sizes of 512 $\times$ 512 $\times$ 391 pixels. In addition, two 2D fluoroscopic (2D-Fluoro) and two 2D-DSA images were acquired from lateral and the AP views, with pixel sizes of \qtyproduct{0.15 x 0.15}{\milli\metre} and image sizes of either 1920 $\times$ 2480 pixels or 2480 $\times$ 1920 pixels. 2D-Fluoro images have lower vessel contrast and display nonvascular anatomical structures and interventional tools, in contrast to 2D-DSA images. All 2D-3D image pairs have ground-truth extrinsics, intrinsics and target points, as described in the work of Mitrovic~\etal~\cite{pernus20133d}. Only the AP views (2D-DSA and 2D-Fluoro) were used to test our fluoro-CT registration approach.

\textit{DeepFluoro}: This dataset includes lower torso CTs from 3 male and 3 female cadaveric specimens. All CTs were resampled to an isotropic spacing of \qtyproduct{1 x 1 x 1}{\milli\metre}. For each CT, 30 synthetic X-ray images were generated using a DRR simulation approach~\cite{grupp2020automatic}, with angular and orbital rotations between \SI{-25}{\degree} and \SI{25}{\degree}, and SIDs set between \SI{90}{\centi\metre} and \SI{125}{\centi\metre}. During simulation, Poisson noise, corresponding to 2000 simulated photons per detector pixel, was introduced to generate more realistic fluoroscopic images. The corresponding intrinsic and extrinsic matrices were used as the ground truth. Pelvic segmentation was performed using image thresholding to extract surface points as 3D target points, and corresponding 2D target points were derived by applying the ground-truth projection matrix to 3D target points.

\begin{figure*}[h!]
\centering
\begin{tabular}{@{} c @{}  c @{}  c @{}  c @{} c @{} c @{}  c @{}  c@{}}

%\begin{tabular}{@{} m{.12\textwidth} @{}  m{.12\textwidth} @{}  m{.12\textwidth} @{}  m{.12\textwidth} @{} m{.12\textwidth} @{}  m{.12\textwidth} @{}  m{.12\textwidth} @{}  m{.12\textwidth}@{} }

%\hline
\includegraphics[width=.11\textwidth]{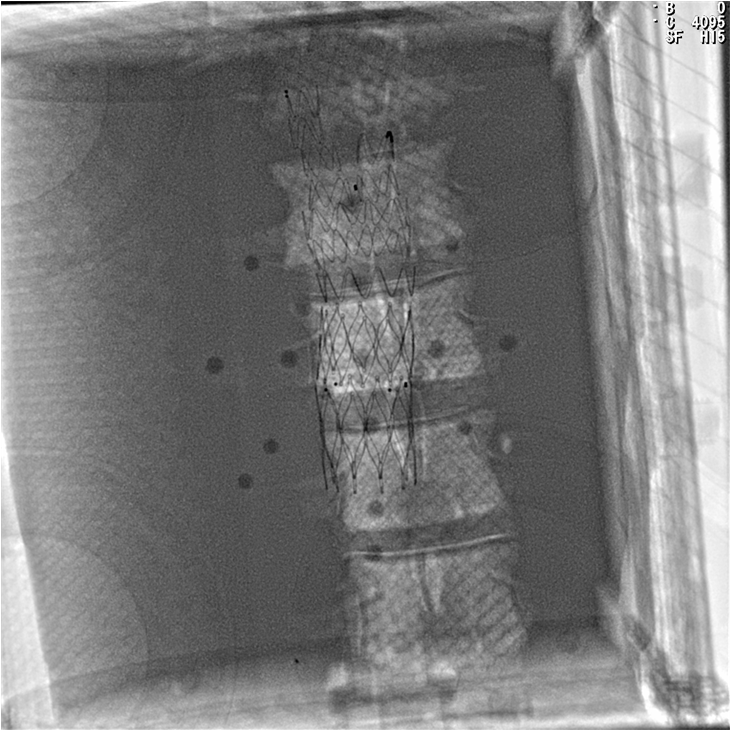}&
\includegraphics[width=.11\textwidth]{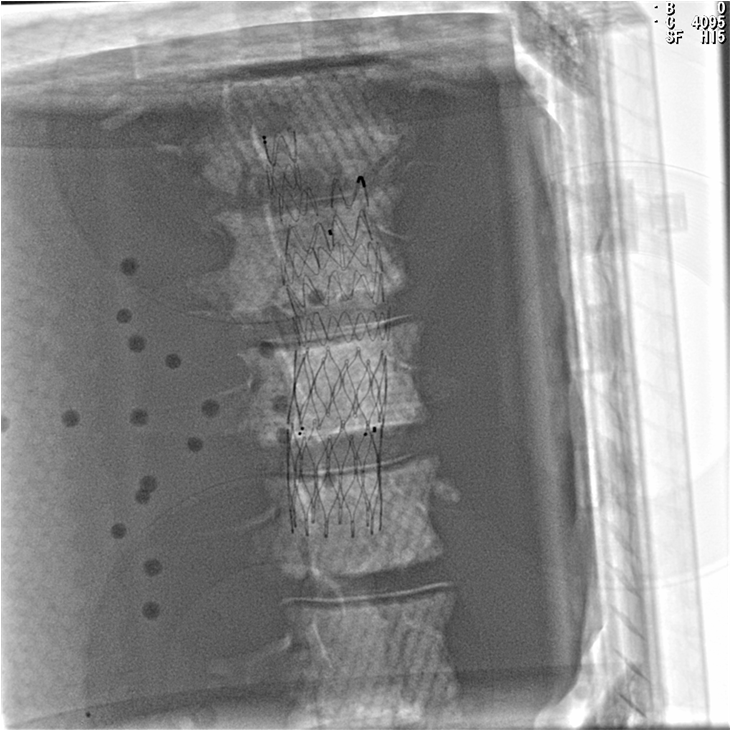}&
\includegraphics[height=.11\textwidth]{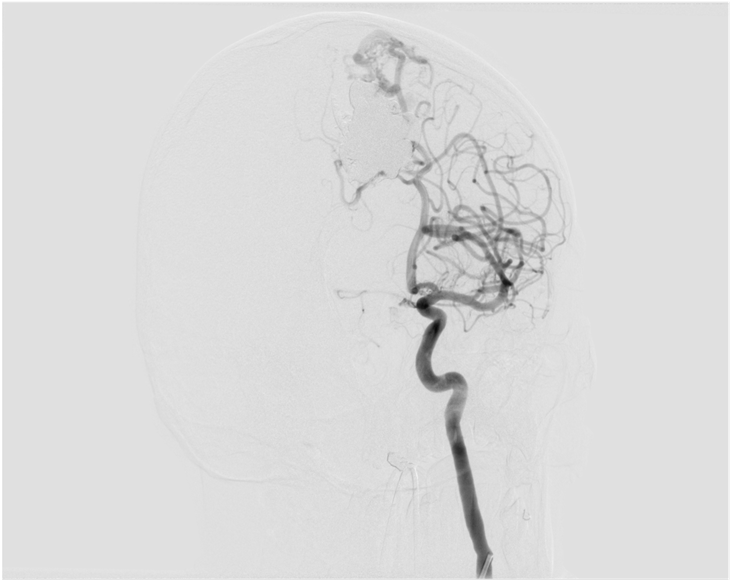}&
\includegraphics[height=.11\textwidth]{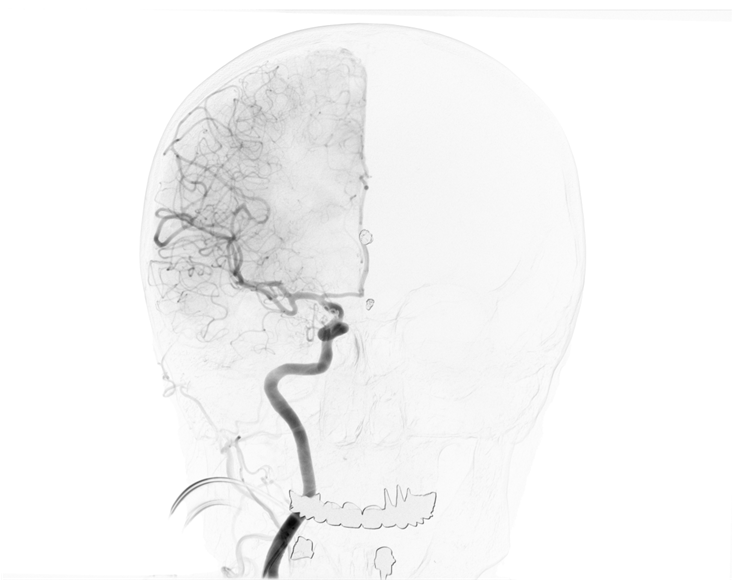}&
\includegraphics[height=.11\textwidth]{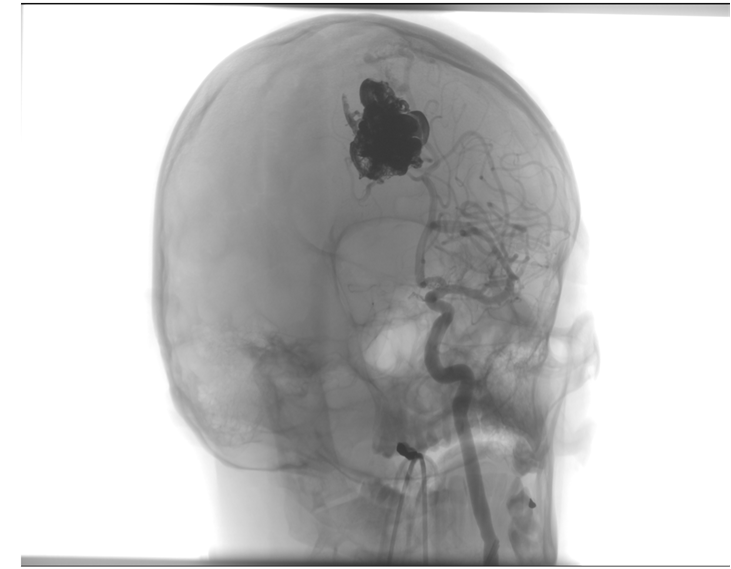}&
\includegraphics[height=.11\textwidth]{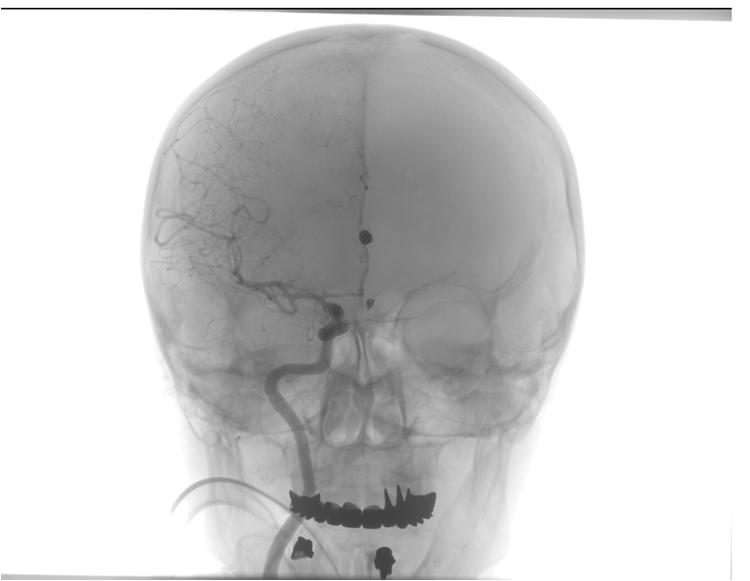}&
\includegraphics[width=.11\textwidth]{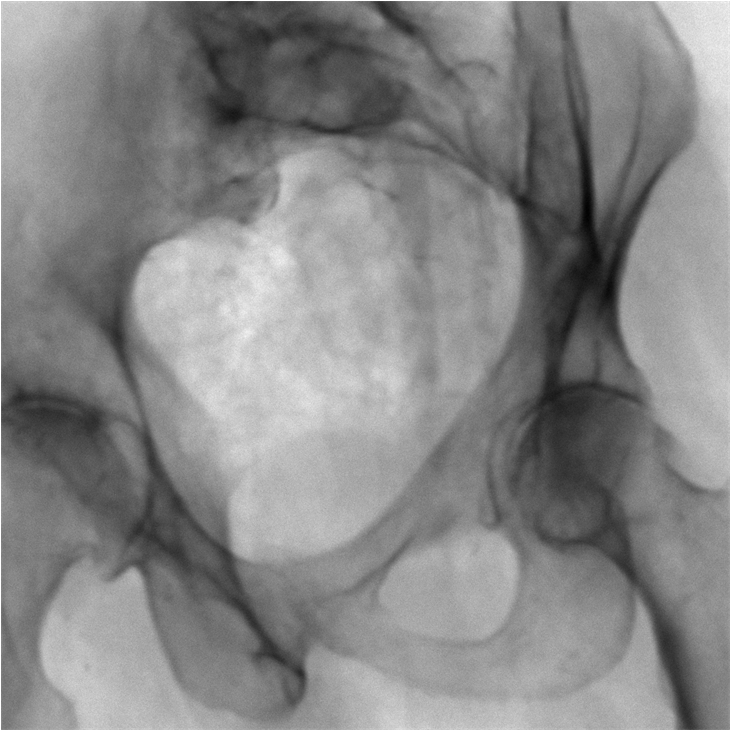}&
\includegraphics[width=.11\textwidth]{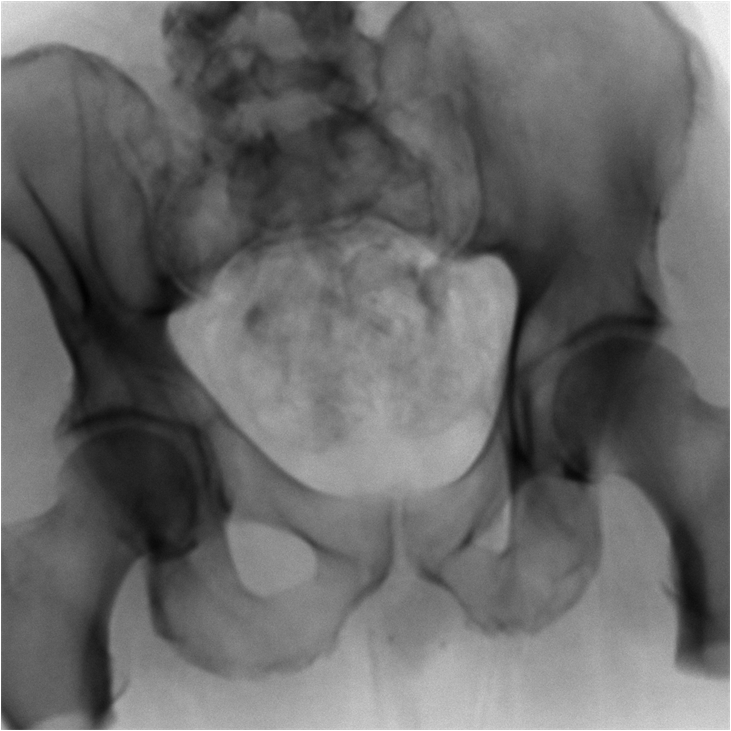}\\

\includegraphics[width=.11\textwidth]{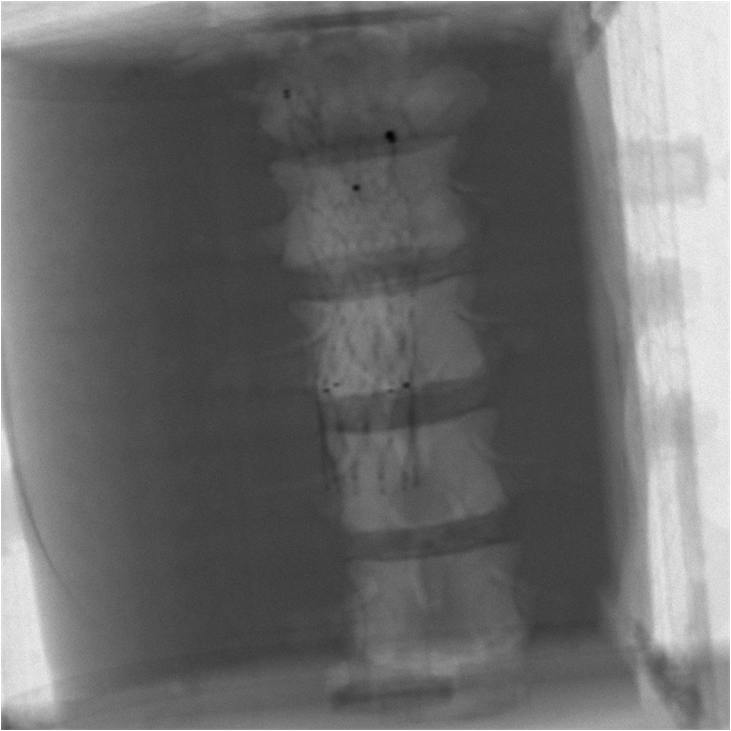}&
\includegraphics[width=.11\textwidth]{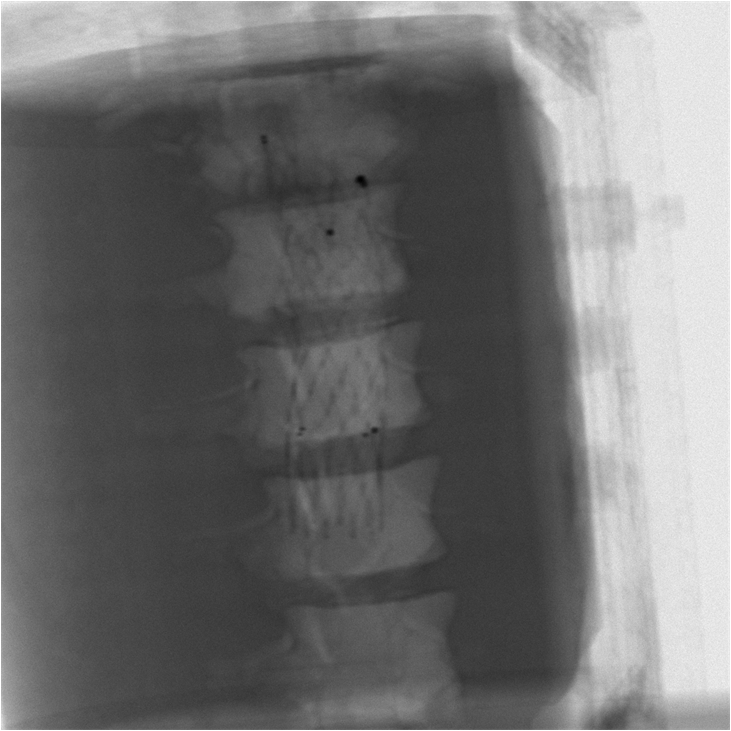}&
\includegraphics[height=.11\textwidth]{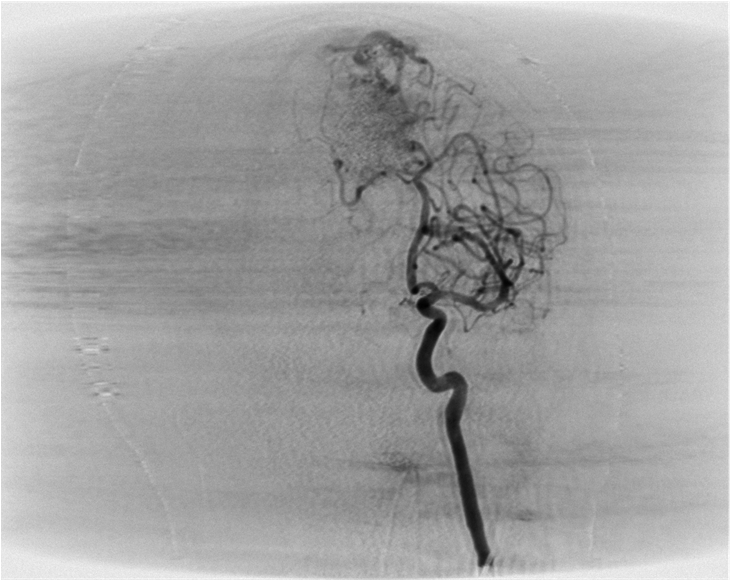}&
\includegraphics[height=.11\textwidth]{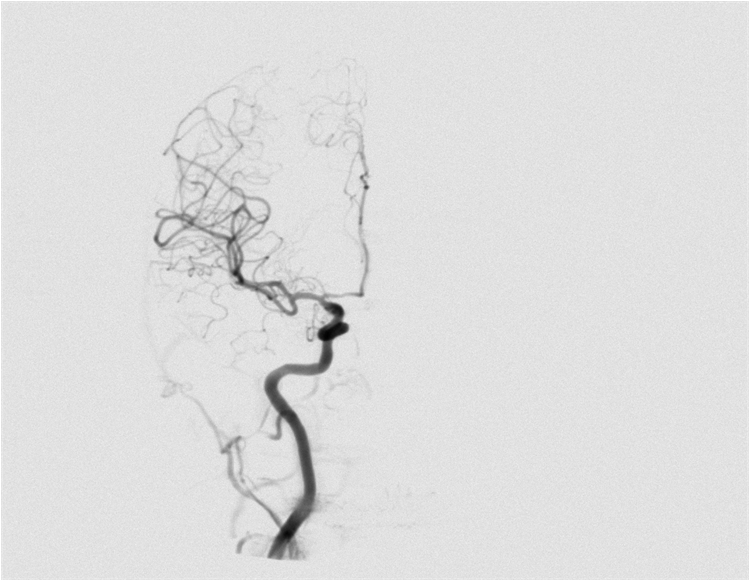}&
\includegraphics[height=.11\textwidth]{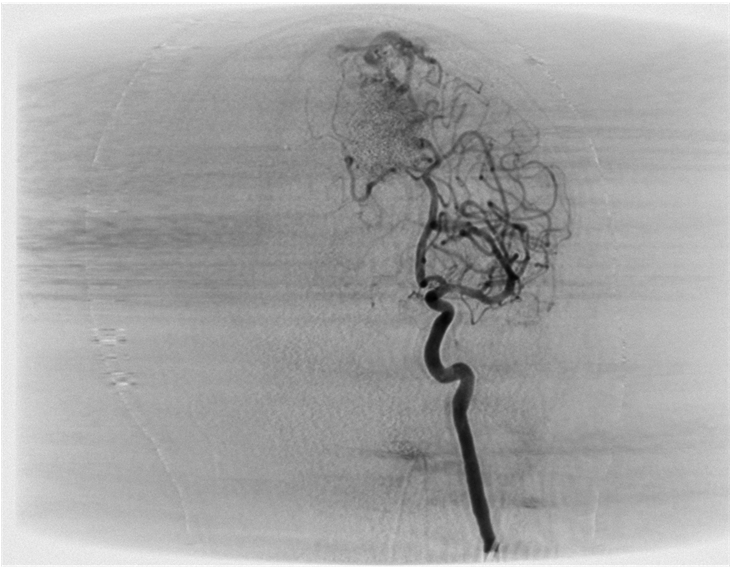}&
\includegraphics[height=.11\textwidth]{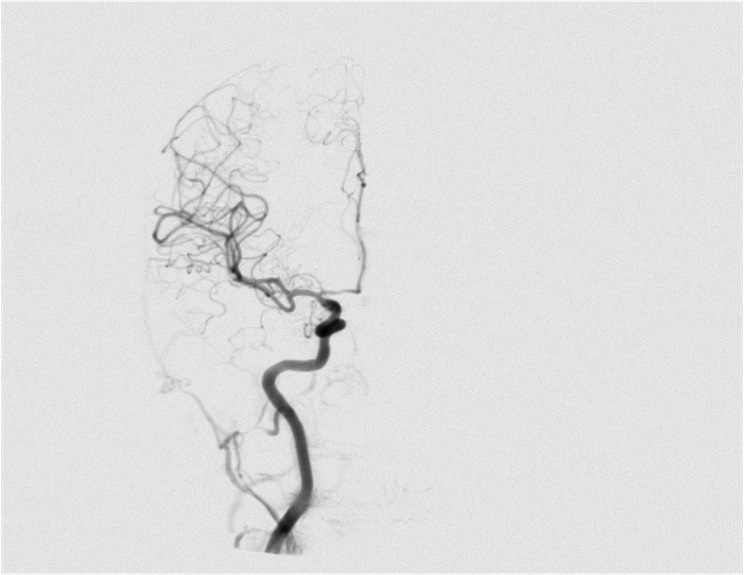}&
\includegraphics[width=.11\textwidth]{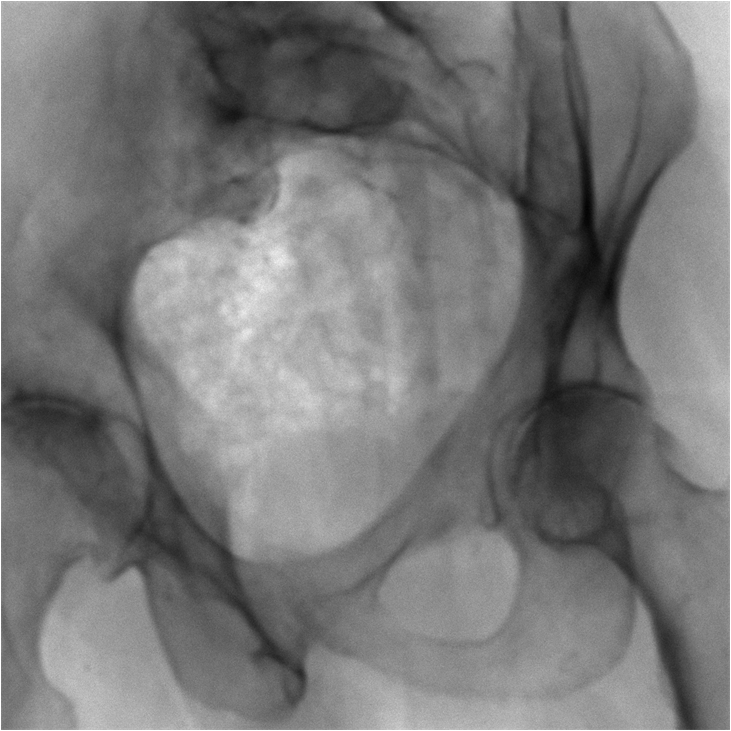}&
\includegraphics[width=.11\textwidth]{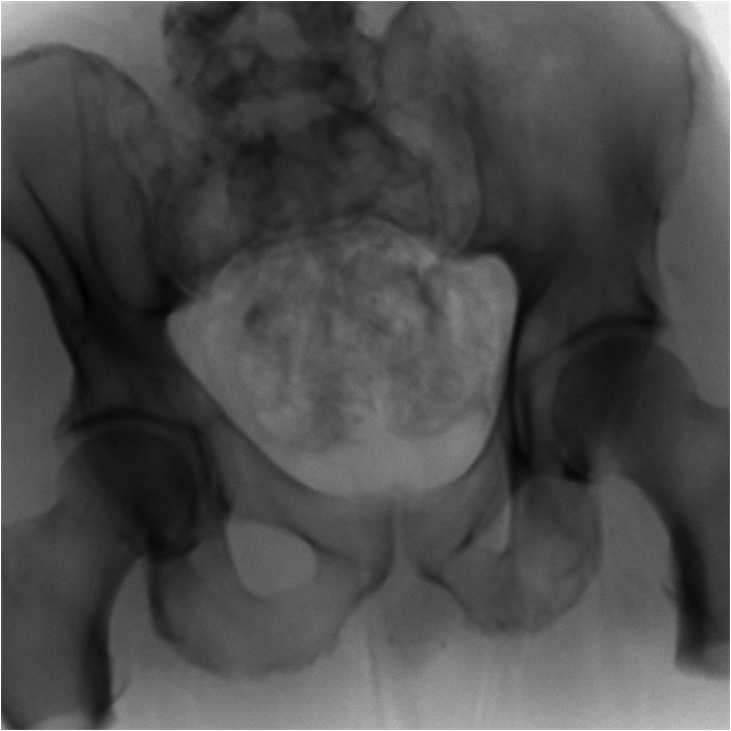}\\

\includegraphics[width=.11\textwidth]{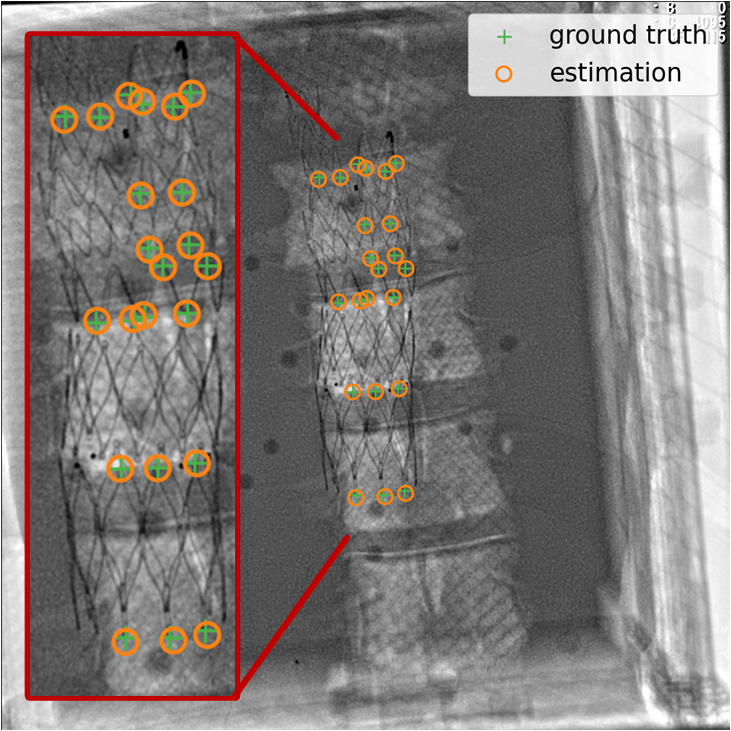}&
\includegraphics[width=.11\textwidth]{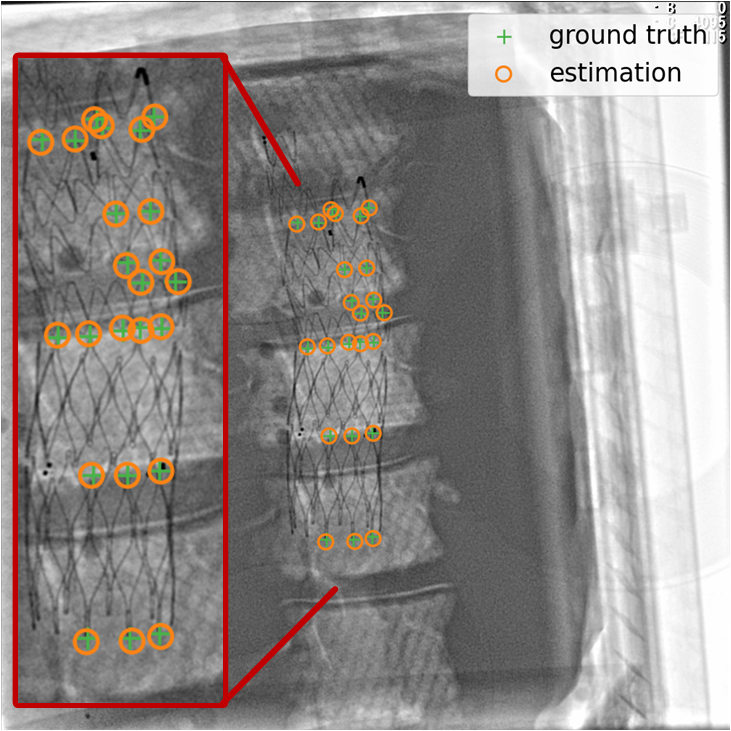}&
\includegraphics[height=.11\textwidth]{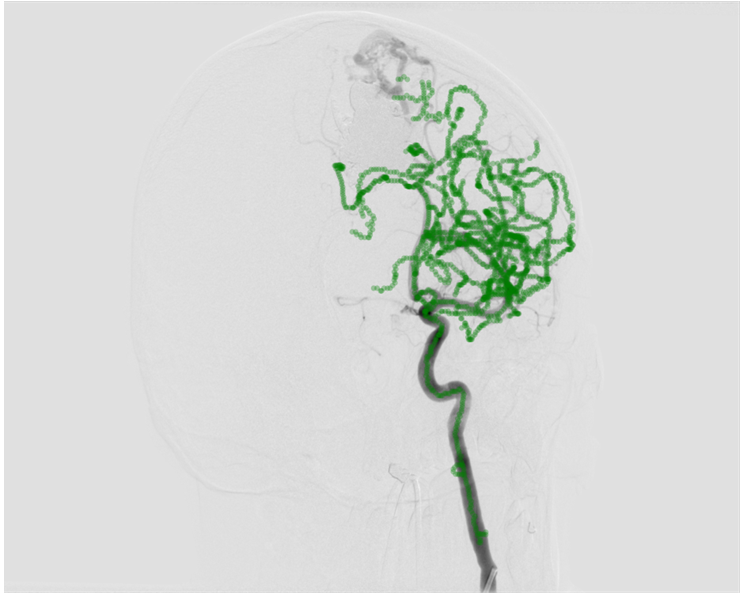}&
\includegraphics[height=.11\textwidth]{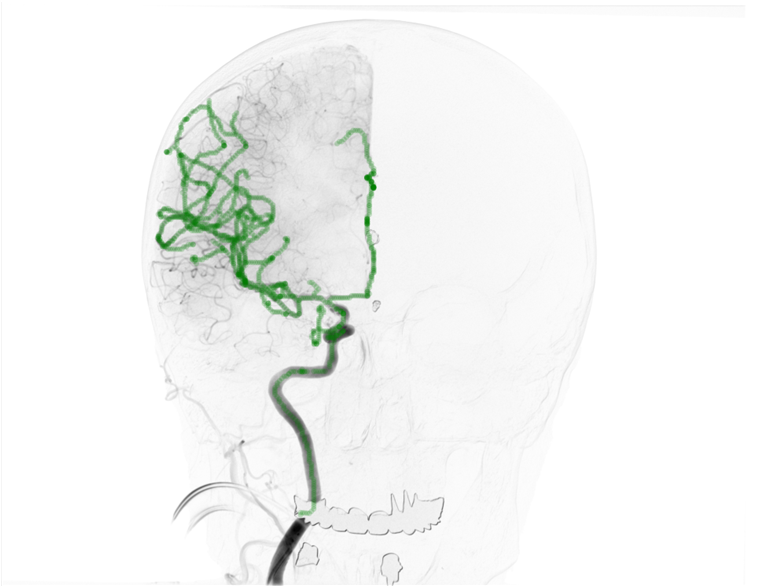}&
\includegraphics[height=.11\textwidth]{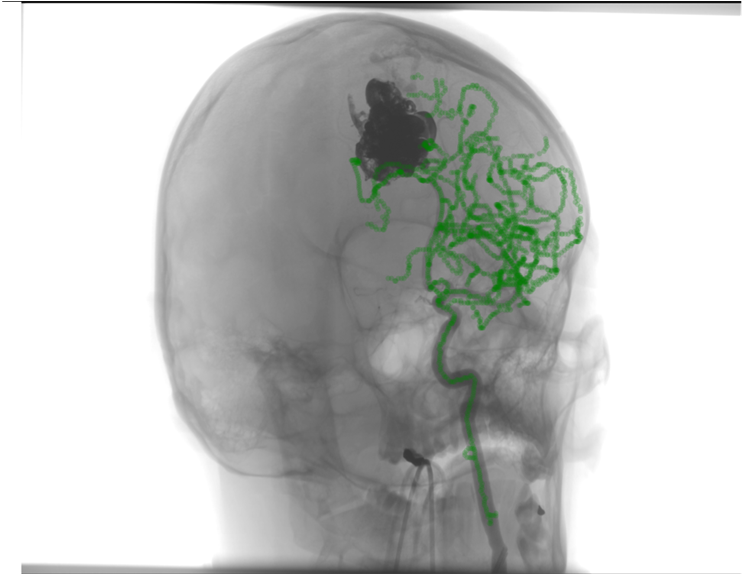}&
\includegraphics[height=.11\textwidth]{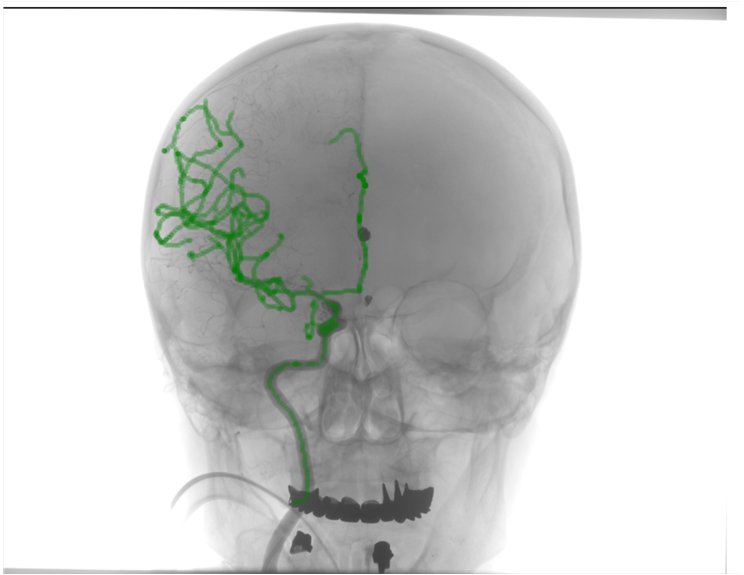}&
\includegraphics[width=.11\textwidth]{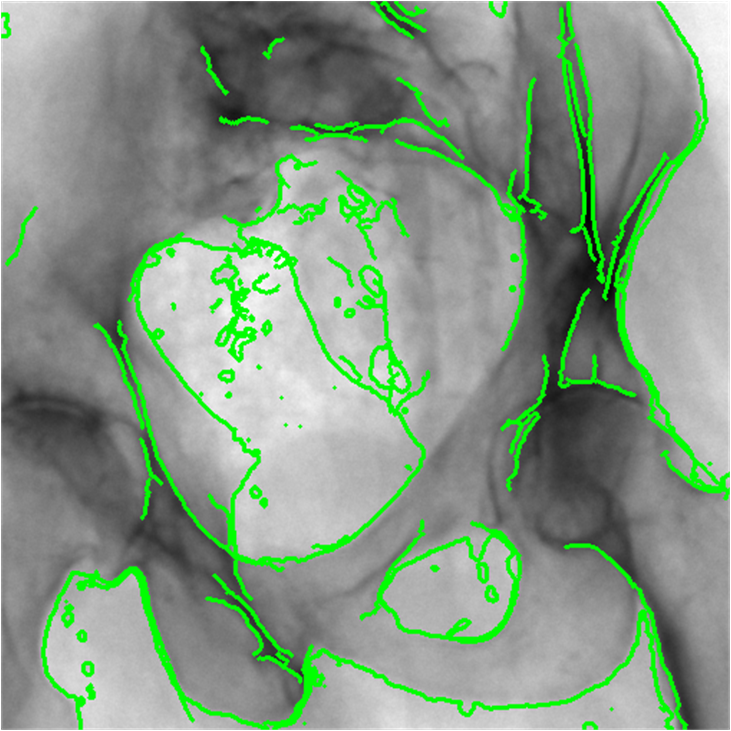}&
\includegraphics[width=.11\textwidth]{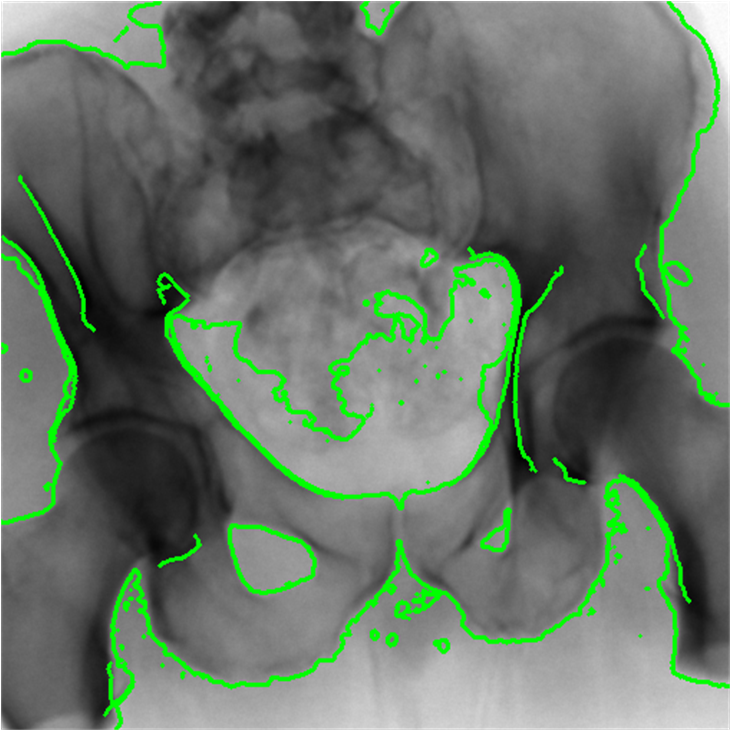}\\

{(A-1)} & {(A-2)} & {(B-1)} & {(B-2)} & {(C-1)} & {(C-2)} & {(D-1)} & {(D-2)}\\
%\multicolumn{3}{c}{C} & \multicolumn{3}{c}{D}\\
%\hline
\end{tabular}
\caption{Results of fluoro-CT registration across various datasets. (For each case, the input fluoroscopy/DSA, the registered image and the input fluoroscopy/DSA overlaid by target landmarks are presented from top to bottom.)}
\label{fig_fluoro_CT_registration_datasets}
\end{figure*}

\begin{figure}[h!]
\centering
\begin{tabular}{@{} c @{}  c @{}  c @{} }
\includegraphics[width=.15\textwidth]{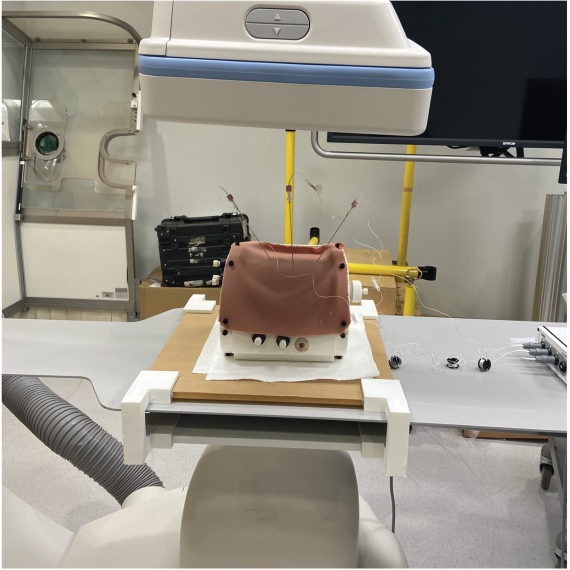}&
\includegraphics[width=.15\textwidth]{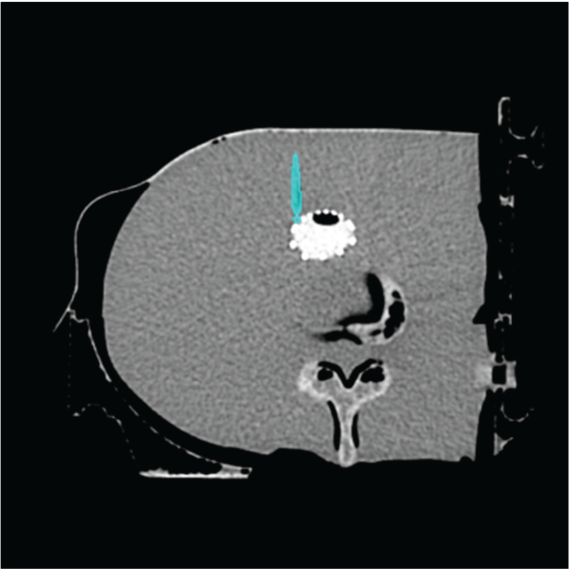}&
\includegraphics[width=.15\textwidth]{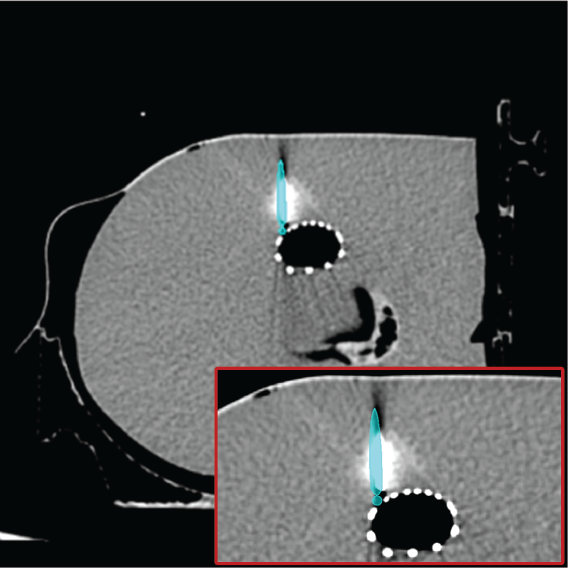}\\

\includegraphics[width=.15\textwidth]{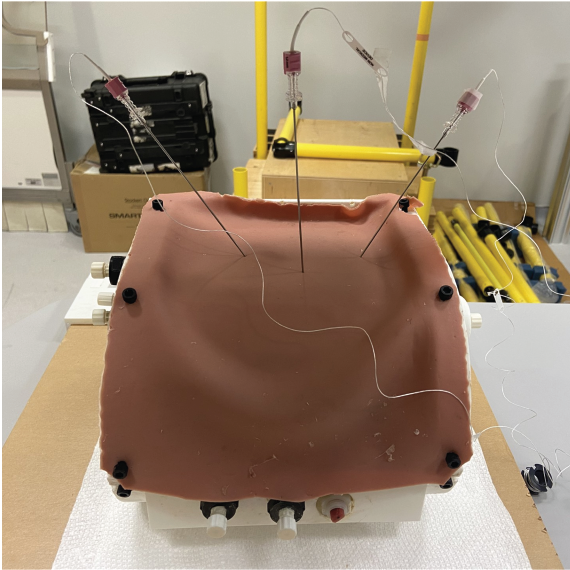}&
\includegraphics[width=.15\textwidth]{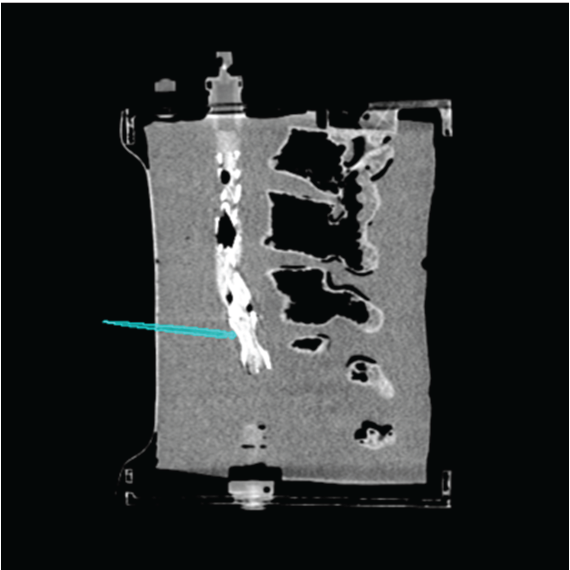}&
\includegraphics[width=.15\textwidth]{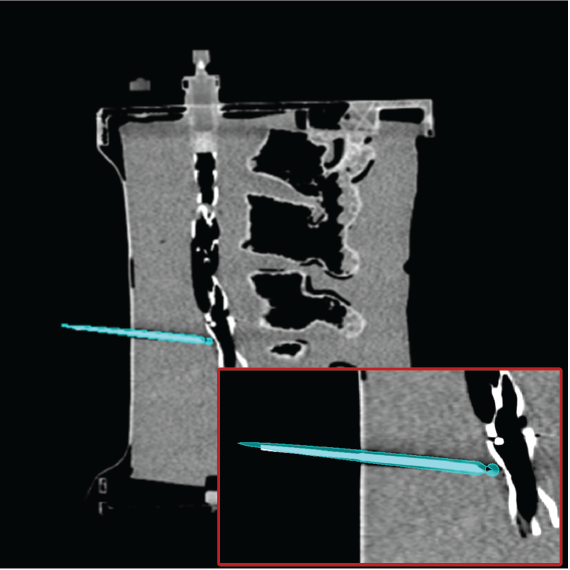}\\

\includegraphics[width=.15\textwidth]{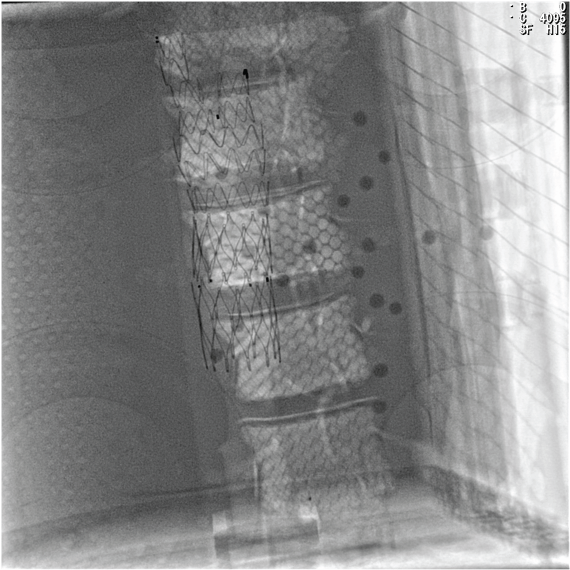}&
\includegraphics[width=.15\textwidth]{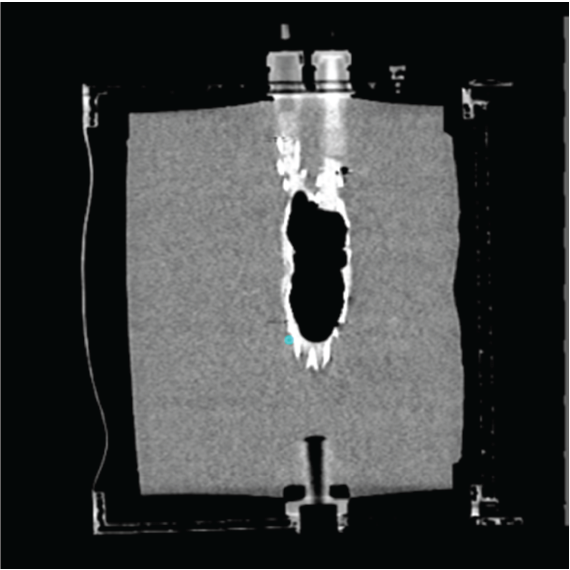}&
\includegraphics[width=.15\textwidth]{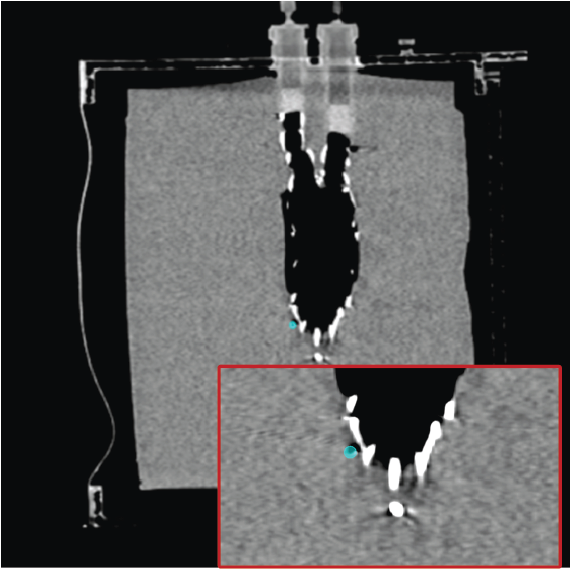}\\

\includegraphics[width=.15\textwidth]{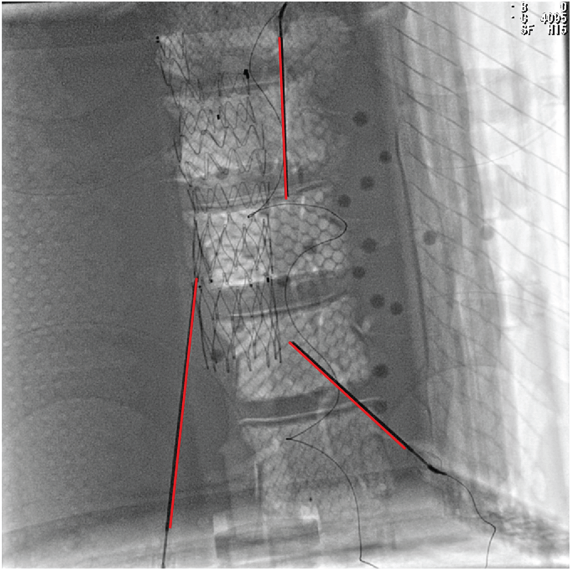} &
\includegraphics[width=.15\textwidth]{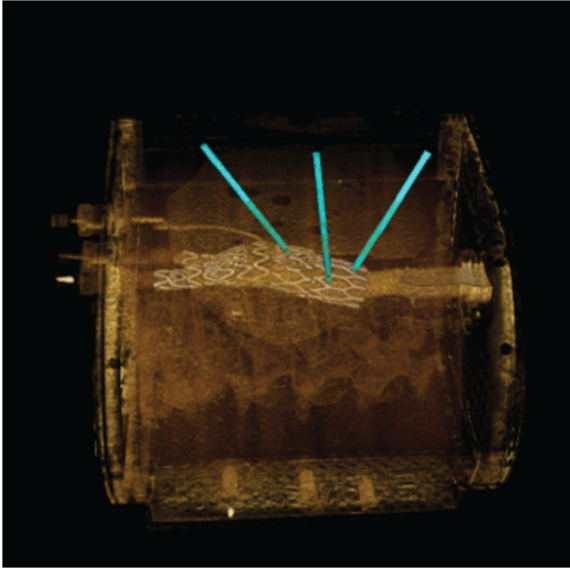} &
\includegraphics[width=.15\textwidth]{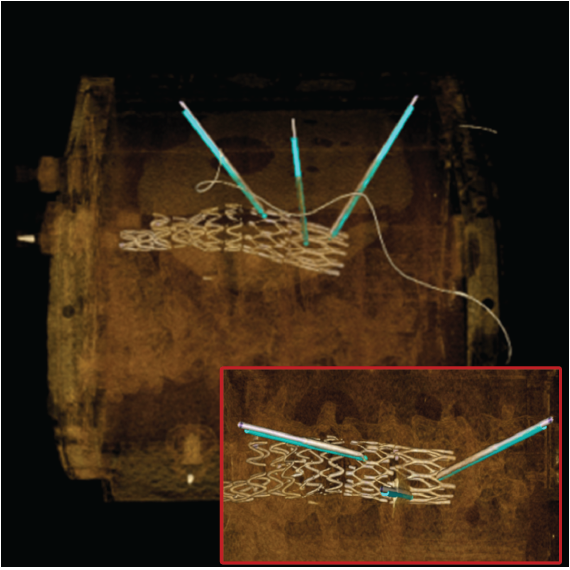}
\\
{(A)} & {(B)} & {(C)}
\end{tabular}
\caption{Setup and results of phantom targeting in 2D and 3D navigation. (A) experimental setup and 2D navigation; (B) 3D navigation and (C) validation of needle insertion in 3D navigation.}
\label{fig_phantom_target_experiments}
\end{figure}

\begin{table*}[h!]
\caption{Registration results on different clinical datasets.}
\centering
\setlength{\tabcolsep}{8pt}
\begin{tabular}{cccccc}
\Xhline{2\arrayrulewidth}
{Dataset} & {\# of patients} &{\# of image pairs} & {mPD(mm)} & {mRPD(mm)} & {mTRE(mm)}\\
\hline
{3D-2D-GS-CA (DSA)} & {10} & {10} & {0.23 $\pm$ 0.10} & {0.15 $\pm$ 0.06} & {0.86 $\pm$ 0.46}\\
{3D-2D-GS-CA (Fluoro)} & {10} & {10} & {0.29 $\pm$ 0.14} & {0.18 $\pm$ 0.08} & {1.41 $\pm$ 0.58}\\
{DeepFluoro} & {6} & {120} & {0.22 $\pm$ 0.04} & {0.15 $\pm$ 0.03} & {0.29 $\pm$ 0.17}\\
\Xhline{2\arrayrulewidth}
\end{tabular}
\label{tab_registration_different_datasets}
\end{table*}

Table~\ref{tab_registration_different_datasets} summarizes the registration errors for the \textit{3D-2D-GS-CA} and \textit{DeepFluoro} datasets. Both datasets achieved similar 2D registration errors, with mPDs $\leq$ \SI{0.3}{\milli\metre} and mRPDs approximately \SI{0.15}{\milli\metre}. For 3D, 2D-Fluoro cases yielded a mean TRE of \SI{1.41}{\milli\metre} $\pm$ \SI{0.58}{\milli\metre}, though 2 out of 10 cases had mTREs exceeding \SI{2.0}{\milli\metre}. Fig.~\ref{fig_fluoro_CT_registration_datasets} shows qualitative results for 8 cases, showing input fluoroscopy/DSA, the registered image, and the input fluoroscopy/DSA with post-registration target points overlaid. All cases demonstrated good alignment, particularly in cases C-1 and C-2, underscoring the approach's effectiveness.
\subsection{Phantom experiments}
A phantom targeting experiment was conducted to evaluate the accuracy of needle insertion. Fig.~\ref{fig_phantom_target_experiments}A shows the experimental setup, including an endoleak phantom, an FG mounting frame, and a Canon Alphenix C-arm (Canon Inc., Tokyo, Japan). Three magnetically tracked needles were inserted into the endoleak phantom. During 2D navigation, the tracked needles (red lines) were projected on the fluoroscopic image, while the actual needles appeared as dark lines. The discrepancies between the two were used to quantify 2D targeting accuracy.

Fig.~\ref{fig_phantom_target_experiments}B illustrates the 3D navigation, including 3D reformations, the rendered contrast-enhanced CT volume, and tracked needles. To validate 3D needle insertion accuracy, we acquired a post-insertion non-contrast CT image (Fig.~\ref{fig_phantom_target_experiments}C). The inserted needles were clearly visible in three orthogonal views, with the 3D view displaying the inserted needles in silver and the tracked needles in cyan. Needle tip and angle errors were calculated by comparing the tracked needles with their segmented counterparts from the CT image. This process was repeated three times, with a total of nine inserted needles. The overall needle insertion errors were \SI{2.65}{\milli\metre} $\pm$ \SI{0.66}{\milli\metre} for the tip, and \SI{1.64}{\degree} $\pm$ \SI{1.40}{\degree} for orientation, as shown in Table~\ref{tab_results_phantoms_targeting}. 

\begin{table}[h!]
\caption{Results of phantom targeting experiments.}
\centering
\setlength{\tabcolsep}{8pt}
\begin{tabular}{ccc}
\Xhline{2\arrayrulewidth}
{Cases} & {Needle tip distance (mm)} &{Needle angle error ($^{\circ}$)}\\
\hline
{1} & {3.46} & {1.76}\\
{2} & {2.87} & {1.02}\\
{3} & {3.95} & {0.13}\\
{4} & {2.03} & {0.57}\\
{5} & {2.20} & {1.62}\\
{6} & {2.52} & {0.52}\\
{7} & {2.11} & {1.32}\\
{8} & {2.40} & {4.14}\\
{9} & {2.32} & {3.72}\\
{mean $\pm$ std} & {2.65 $\pm$ 0.66} & {1.64 $\pm$ 1.40}\\
\Xhline{2\arrayrulewidth}
\end{tabular}
\label{tab_results_phantoms_targeting}
\end{table}

As shown in Table~\ref{tab_runtime_workflow}, our deep decomposition model required a mean of \SI{0.07}{\second} to remove metal-induced artifacts from X-ray images on an RTX 2080Ti card. The parameters of the mechanically driven C-arm were automatically recognized using an OCR technique with a mean runtime of \SI{0.15}{\second} for a single-channel input. In 2D navigation, surgical instruments were virtually augmented in real time using tracking data. For 3D navigation, fluoro-CT registration required a mean of \SI{1.97}{\second}, with fluoroscopic image dimensions of 1017 $\times$ 1017 pixels and pre-operative CT sizes of 365 $\times$ 280 $\times$ 797 voxels. After registration, the instruments can be tracked in 3D for real-time guidance. 

\begin{table}[h!]
\caption{Runtime of each step during surgical workflows.}
\centering
\setlength{\tabcolsep}{8pt}
\begin{tabular}{cc|c}
\Xhline{2\arrayrulewidth}
\multicolumn{2}{c|}{Surgical workflows} & {Runtime/frame rate}\\
\hline
\multirow{3}{*}{Input} & {Frames grabbing} & {30 fps}\\
{} & {Fluoro decomposition} & {$\sim$0.07 s}\\
{} & {C-arm parameters capture} & {$\sim$0.15 s}\\
\hline
{2D Navigation} & {Virtual roadmap of instruments} & {30 fps} \\
\hline
\multirow{2}{*}{3D Navigation} & {Fluoro-CT registration} & {$\sim$1.97 s}\\
{} & {Tracking of instruments} & {30 fps}\\
\hline

\Xhline{2\arrayrulewidth}
\end{tabular}
\label{tab_runtime_workflow}
\end{table}

\section{Discussion}
A prototype magnetic tracker with radio\-lucent magnetic field generator, developed by Northern Digital Inc., was used in this study, marking the first step toward integrating a magnetic tracking (MT) system into fluoroscopy-guided procedures. Our proposed MT-based workflow demonstrated its efficacy in facilitating fluoroscopy guidance across three aspects: hardware design, software deployment and phantom experiments. 
\begin{enumerate}
    \item Hardware design. The two-layer FG mounting frame ensures smooth installation on various C-arm systems and optimal MT accuracy at the surgical site. In contrast to other configurations (\eg, above or adjacent to the patient but not on the surgical table), our integrated FG mounting frame does not limit the C-arm flexibility during procedures. For instance, the cone-beam CT acquisition, a process that needs a wide range rotation of the C-arm, can be performed without hindrance. In addition, the incorporation of external fiducials demonstrated its advantages in estimating the C-arm pose, thereby improving the efficacy of the fluoro-CT registration. Although aluminum fiducials result in external metal-induced artifacts in fluoroscopic images, our deep decomposition model successfully removed these artifacts without affecting guidance. This step requires no additional computational burden because our decomposition model initially tends to remove the limited FG-induced artifacts, including the traces of FG coils and electronic components. 
    \item Software deployment. Our OpenIGTLink-based server-client architecture enabled seamless streaming of multi-modal inputs into fluoroscopy-guided workflows. Combing 3D Slicer client with readily available modules, it is straightforward and efficient to build procedure-specific workflows, including data collection, algorithm processing, visualization and others. Additionally, the Plus Server provides high-resolution timestamps, enabling synchronization of live data (\eg, fluoroscopic images and instrument tracking) for guidance. Our framework is extensible, allowing for the integration of additional modalities without disrupting existing workflows.
    \item Phantom experiments. We conducted an endoleak phantom targeting experiment to assess the clinical feasibility of our approach, focusing on needle insertion accuracy and workflow runtime. Needle insertion accuracy was consistent, with errors ranging from \SI{2}{\milli\metre} to \SI{3}{\milli\metre} across nine tasks, which is suitable for most clinical procedures.  Fluoroscopic image confirmation during procedures further supports the potential for precise 3D navigation ($<$ \SI{2}{\milli\metre}) in surgical tasks. In terms of workflow runtime, artifact-free fluoroscopic images were generated at approximately 14 fps ($\sim$\SI{0.07}{\second}), which is deemed clinically acceptable for subsequent navigation tasks. For mechanically driven systems, our proposed workflow required $\sim$\SI{0.15}{\second} ($\sim$6 fps) to run the OCR approach for updating the C-arm parameter, with negligible impact on 3D navigation. Meanwhile, 2D navigation continues to execute in real time.
\end{enumerate}
Our proposed 3D navigation can be robustly applied to different C-arm configurations. One critical technical component in 3D navigation is achieving effective fluoro-CT registration, which depends on robust estimation of the C-arm pose. Our results (Fig.~\ref{fig_mPD_movement_fiducials}B) indicate that the C-arm pose can be accurately estimated with as few as six fiducials, making our solution applicable to a wide range of clinical scenarios. However, due to the projective nature of X-ray images, some fiducials may have limited visibility or fall outside of the imaging field of view. Thus, it remains possible that the number of captured fiducials may fall short of this minimal requirement, particularly in lateral view. 3D navigation aims at establishing the relative relationship between the patient and the FG mounting frame, (i.e., $T_{FG}^{patient} = (T_{patient}^{source})^{-1}*T_{FG}^{source}$). In theory, this step does not necessitate registration at a specific C-arm pose, such as the lateral view. Any other views containing enough fiducials ($>$ 5) can be used to determine the transformation $T_{FG}^{patient}$. As a result, 3D navigation remains effective even in challenging cases with limited fiducial visibility.

Although a “3D roadmap” feature is available in high-end, mechanically driven C-arms, our proposed 3D navigation offers advantages. The “3D roadmap” overlays a pre-acquired 3D image, typically from a cone-beam CT, onto live fluoroscopic images to assist with guidance. However, if the patient or surgical table moves during the procedure, manual adjustments to the registration are often required, which can be time-consuming and prone to registration failure. In contrast, our 3D navigation uses fiducials attached to the FG mounting frame to estimate the current C-arm pose, enabling fluoro-CT registration that is unaffected by surgical table movement. In addition, as shown in Fig.~\ref{fig_mPD_movement_fiducials}A, our approach can accommodate patient movements within a \qtyproduct{30x30}{\milli\metre} region. Beyond this range, the validity of our approach may be compromised. Fortunately, \qtyproduct{30 x 30}{\milli\metre} is clinically reasonable working space in terms of any patient movement.

When tested on the endoleak phantom, our approach achieved a mPD of approximately \SI{0.7}{\milli\metre}. However, applying the registration to two public datasets resulted in a significantly lower mPD of around \SI{0.25}{\milli\metre} - a factor of three lower than in the phantom test. In our registration workflow, C-arm intrinsic parameters, including pixel spacing, image size and focal length (\ie, SID), were treated as known \textit{a priori}. Due to the inherent projective nature of X-ray imaging, any variations in intrinsics may be compensated by adjusting the extrinsics to achieve optimal alignment when running our registration algorithm. In the endoleak phantom experiment, we used C-arm monitor readings (SID) to calculate the intrinsic parameters. However, due to the C-arm gantry flexing under gravity as it rotates~\cite{jain2005c}, the actual SID may be different from the C-arm monitor reading, which primarily contributes to the lower fluoro-CT registration accuracy. In the clinical setting, it is challenging to obtain calibrated C-arm intrinsics (for example, the two public datasets). Therefore, a mPD of approximately \SI{0.7}{\milli\metre} may be more representative of expected performance during procedures. 

\section{Conclusion}
To facilitate the seamless integration of MT into fluoroscopy-guided interventions, we designed a two-layer mounting frame incorporating a tabletop radiolucent FG, for attachment to the surgical table and optimal tracking accuracy, making it applicable to a wide range of surgical sites under fluoroscopy guidance. Without interfering with the conventional workflow, we proposed a strategy of adding external fiducials to address key technical challenges, including C-arm pose estimation and robust fluoro-CT registration. To provide depth information, we complemented the surgical workflow with 3D navigation, as well as enhanced 2D navigation. Results demonstrated the effectiveness and clinical applicability of this MT-assisted solution. 

\section*{Acknowledgment}
We gratefully acknowledge the support of NVIDIA Corporation with the donation of the Quadro RTX 6000 used for this research.

\bibliographystyle{IEEEtran}
\bibliography{biblio}

% Generated by IEEEtran.bst, version: 1.14 (2015/08/26)
\begin{thebibliography}{10}
\providecommand{\url}[1]{#1}
\csname url@samestyle\endcsname
\providecommand{\newblock}{\relax}
\providecommand{\bibinfo}[2]{#2}
\providecommand{\BIBentrySTDinterwordspacing}{\spaceskip=0pt\relax}
\providecommand{\BIBentryALTinterwordstretchfactor}{4}
\providecommand{\BIBentryALTinterwordspacing}{\spaceskip=\fontdimen2\font plus
\BIBentryALTinterwordstretchfactor\fontdimen3\font minus \fontdimen4\font\relax}
\providecommand{\BIBforeignlanguage}[2]{{%
\expandafter\ifx\csname l@#1\endcsname\relax
\typeout{** WARNING: IEEEtran.bst: No hyphenation pattern has been}%
\typeout{** loaded for the language `#1'. Using the pattern for}%
\typeout{** the default language instead.}%
\else
\language=\csname l@#1\endcsname
\fi
#2}}
\providecommand{\BIBdecl}{\relax}
\BIBdecl

\bibitem{merloz2007fluoroscopy}
P.~Merloz, J.~Troccaz, H.~Vouaillat, C.~Vasile, J.~Tonetti, A.~Eid, and S.~Plaweski, ``Fluoroscopy-based navigation system in spine surgery,'' \emph{Proceedings of the Institution of Mechanical Engineers, Part H: Journal of Engineering in Medicine}, vol. 221, no.~7, pp. 813--820, 2007.

\bibitem{cazzato2020spinal}
R.~L. Cazzato, P.~Auloge, P.~De~Marini, E.~Boatta, G.~Koch, D.~Dalili, P.~P. Rao, J.~Garnon, and A.~Gangi, ``Spinal tumor ablation: indications, techniques, and clinical management,'' \emph{Techniques in Vascular and Interventional Radiology}, vol.~23, no.~2, p. 100677, 2020.

\bibitem{nijkamp2019prospective}
J.~Nijkamp, K.~F. Kuhlmann, O.~Ivashchenko, B.~Pouw, N.~Hoetjes, M.~A. Lindenberg, A.~G. Aalbers, G.~L. Beets, F.~van Coevorden, N.~KoK \emph{et~al.}, ``Prospective study on image-guided navigation surgery for pelvic malignancies,'' \emph{Journal of Surgical Oncology}, vol. 119, no.~4, pp. 510--517, 2019.

\bibitem{gao2022fluoroscopy}
C.~Gao, H.~Phalen, A.~Margalit, J.~H. Ma, P.-C. Ku, M.~Unberath, R.~H. Taylor, A.~Jain, and M.~Armand, ``Fluoroscopy-guided robotic system for transforaminal lumbar epidural injections,'' \emph{IEEE transactions on medical robotics and bionics}, vol.~4, no.~4, pp. 901--909, 2022.

\bibitem{bakhtiarinejad2023surgical}
M.~Bakhtiarinejad, C.~Gao, A.~Farvardin, G.~Zhu, Y.~Wang, J.~K. Oni, R.~H. Taylor, and M.~Armand, ``A surgical robotic system for osteoporotic hip augmentation: System development and experimental evaluation,'' \emph{IEEE transactions on medical robotics and bionics}, vol.~5, no.~1, pp. 18--29, 2023.

\bibitem{ramadani2022survey}
A.~Ramadani, M.~Bui, T.~Wendler, H.~Schunkert, P.~Ewert, and N.~Navab, ``A survey of catheter tracking concepts and methodologies,'' \emph{Medical image analysis}, vol.~82, p. 102584, 2022.

\bibitem{mccutcheon2004frameless}
I.~E. McCutcheon, R.~S. Kitagawa, P.~F. Demasi, B.~K. Law, and K.~E. Friend, ``Frameless stereotactic navigation in transsphenoidal surgery: comparison with fluoroscopy,'' \emph{Stereotactic and functional neurosurgery}, vol.~82, no.~1, pp. 43--48, 2004.

\bibitem{hummel2002evaluation}
J.~Hummel, M.~Figl, C.~Kollmann, H.~Bergmann, and W.~Birkfellner, ``Evaluation of a miniature electromagnetic position tracker,'' \emph{Medical physics}, vol.~29, no.~10, pp. 2205--2212, 2002.

\bibitem{yaniv2006fluoroscopy}
Z.~Yaniv and K.~Cleary, ``Fluoroscopy based accuracy assessment of electromagnetic tracking,'' in \emph{Medical Imaging 2006: Visualization, Image-Guided Procedures, and Display}, vol. 6141.\hskip 1em plus 0.5em minus 0.4em\relax SPIE, 2006, pp. 168--174.

\bibitem{lugez2015electromagnetic}
E.~Lugez, H.~Sadjadi, D.~R. Pichora, R.~E. Ellis, S.~G. Akl, and G.~Fichtinger, ``Electromagnetic tracking in surgical and interventional environments: usability study,'' \emph{International journal of computer assisted radiology and surgery}, vol.~10, pp. 253--262, 2015.

\bibitem{xu2023surgical}
D.~Xu, X.~Ma, L.~Xie, C.~Zhou, and B.~Kong, ``Surgical precision and efficiency of a novel electromagnetic system compared to a robot-assisted system in percutaneous pedicle screw placement of endo-lif,'' \emph{Global Spine Journal}, vol.~13, no.~5, pp. 1243--1251, 2023.

\bibitem{yoo2013electromagnetic}
J.~Yoo, S.~Schafer, A.~Uneri, Y.~Otake, A.~J. Khanna, and J.~H. Siewerdsen, ``An electromagnetic “tracker-in-table” configuration for x-ray fluoroscopy and cone-beam ct-guided surgery,'' \emph{International journal of computer assisted radiology and surgery}, vol.~8, pp. 1--13, 2013.

\bibitem{xia2023x}
W.~Xia, S.~Xing, U.~Jarayathne, U.~Pardasani, T.~Peters, and E.~Chen, ``X-ray image decomposition for improved magnetic navigation,'' \emph{International Journal of Computer Assisted Radiology and Surgery}, vol.~18, no.~7, pp. 1225--1233, 2023.

\bibitem{o2021radiolucent}
K.~O’Donoghue, H.~A. Jaeger, and P.~Cantillon-Murphy, ``A radiolucent electromagnetic tracking system for use with intraoperative x-ray imaging,'' \emph{Sensors}, vol.~21, no.~10, p. 3357, 2021.

\bibitem{stockle2007image}
U.~St{\"o}ckle, K.~Schaser, and B.~K{\"o}nig, ``Image guidance in pelvic and acetabular surgery—expectations, success and limitations,'' \emph{Injury}, vol.~38, no.~4, pp. 450--462, 2007.

\bibitem{cleary2010image}
K.~Cleary and T.~M. Peters, ``Image-guided interventions: technology review and clinical applications,'' \emph{Annual review of biomedical engineering}, vol.~12, no.~1, pp. 119--142, 2010.

\bibitem{unberath2021impact}
M.~Unberath, C.~Gao, Y.~Hu, M.~Judish, R.~H. Taylor, M.~Armand, and R.~Grupp, ``The impact of machine learning on 2d/3d registration for image-guided interventions: A systematic review and perspective,'' \emph{Frontiers in Robotics and AI}, vol.~8, p. 716007, 2021.

\bibitem{markelj2012review}
P.~Markelj, D.~Toma{\v{z}}evi{\v{c}}, B.~Likar, and F.~Pernu{\v{s}}, ``A review of 3d/2d registration methods for image-guided interventions,'' \emph{Medical image analysis}, vol.~16, no.~3, pp. 642--661, 2012.

\bibitem{esteban2019towards}
J.~Esteban, M.~Grimm, M.~Unberath, G.~Zahnd, and N.~Navab, ``Towards fully automatic x-ray to ct registration,'' in \emph{Medical Image Computing and Computer Assisted Intervention--MICCAI 2019: 22nd International Conference, Shenzhen, China, October 13--17, 2019, Proceedings, Part VI 22}.\hskip 1em plus 0.5em minus 0.4em\relax Springer, 2019, pp. 631--639.

\bibitem{franz2014electromagnetic}
A.~M. Franz, T.~Haidegger, W.~Birkfellner, K.~Cleary, T.~M. Peters, and L.~Maier-Hein, ``Electromagnetic tracking in medicine—a review of technology, validation, and applications,'' \emph{IEEE transactions on medical imaging}, vol.~33, no.~8, pp. 1702--1725, 2014.

\bibitem{zou2020deep}
Z.~Zou, S.~Lei, T.~Shi, Z.~Shi, and J.~Ye, ``Deep adversarial decomposition: A unified framework for separating superimposed images,'' in \emph{Proceedings of the IEEE/CVF conference on computer vision and pattern recognition}, 2020, pp. 12\,806--12\,816.

\bibitem{klema1980singular}
V.~Klema and A.~Laub, ``The singular value decomposition: Its computation and some applications,'' \emph{IEEE Transactions on automatic control}, vol.~25, no.~2, pp. 164--176, 1980.

\bibitem{hartley2003multiple}
R.~Hartley and A.~Zisserman, \emph{Multiple view geometry in computer vision}.\hskip 1em plus 0.5em minus 0.4em\relax Cambridge university press, 2003.

\bibitem{grupp2020automatic}
R.~B. Grupp, M.~Unberath, C.~Gao, R.~A. Hegeman, R.~J. Murphy, C.~P. Alexander, Y.~Otake, B.~A. McArthur, M.~Armand, and R.~H. Taylor, ``Automatic annotation of hip anatomy in fluoroscopy for robust and efficient 2d/3d registration,'' \emph{International journal of computer assisted radiology and surgery}, vol.~15, pp. 759--769, 2020.

\bibitem{grupp2019pose}
R.~B. Grupp, R.~A. Hegeman, R.~J. Murphy, C.~P. Alexander, Y.~Otake, B.~A. McArthur, M.~Armand, and R.~H. Taylor, ``Pose estimation of periacetabular osteotomy fragments with intraoperative x-ray navigation,'' \emph{IEEE transactions on biomedical engineering}, vol.~67, no.~2, pp. 441--452, 2019.

\bibitem{hansen2001completely}
N.~Hansen and A.~Ostermeier, ``Completely derandomized self-adaptation in evolution strategies,'' \emph{Evolutionary computation}, vol.~9, no.~2, pp. 159--195, 2001.

\bibitem{powell2009bobyqa}
M.~J. Powell \emph{et~al.}, ``The bobyqa algorithm for bound constrained optimization without derivatives,'' \emph{Cambridge NA Report NA2009/06, University of Cambridge, Cambridge}, vol.~26, pp. 26--46, 2009.

\bibitem{van2005standardized}
E.~B. Van~de Kraats, G.~P. Penney, D.~Tomazevic, T.~Van~Walsum, and W.~J. Niessen, ``Standardized evaluation methodology for 2-d-3-d registration,'' \emph{IEEE transactions on medical imaging}, vol.~24, no.~9, pp. 1177--1189, 2005.

\bibitem{dunnett1955multiple}
C.~W. Dunnett, ``A multiple comparison procedure for comparing several treatments with a control,'' \emph{Journal of the American Statistical Association}, vol.~50, no. 272, pp. 1096--1121, 1955.

\bibitem{chen2015management}
J.~Chen and S.~W. Stavropoulos, ``Management of endoleaks,'' in \emph{Seminars in interventional radiology}, vol.~32, no.~03.\hskip 1em plus 0.5em minus 0.4em\relax Thieme Medical Publishers, 2015, pp. 259--264.

\bibitem{pernus20133d}
F.~Pernus \emph{et~al.}, ``3d-2d registration of cerebral angiograms: A method and evaluation on clinical images,'' \emph{IEEE transactions on medical imaging}, vol.~32, no.~8, pp. 1550--1563, 2013.

\bibitem{jain2005c}
A.~Jain, R.~Kon, Y.~Zhou, and G.~Fichtinger, ``C-arm calibration--is it really necessary?'' in \emph{International Conference on Medical Image Computing and Computer-Assisted Intervention}.\hskip 1em plus 0.5em minus 0.4em\relax Springer, 2005, pp. 639--646.

\end{thebibliography}

\end{document}